\documentclass[
  journal=pasa,
  manuscript=research-paper, 
  year=2020,
  volume=37,
]{cup-journal}

\usepackage{microtype,siunitx,booktabs}
\usepackage{caption}
\usepackage{subcaption}
\usepackage{verbatim}  
\usepackage{float}
\usepackage{xcolor}     
\usepackage{siunitx}
\usepackage{rotating}
\usepackage{amsfonts}
\usepackage[hidelinks]{hyperref}
\usepackage{cleveref}
\crefname{section}{§}{§§}
\Crefname{section}{§}{§§}
\usepackage{longtable}
\usepackage{natbib}
\usepackage{verbatim}
\usepackage{lscape}
\usepackage{pdflscape}
\usepackage{graphics}
\usepackage{epsfig}
\usepackage{multirow}
\usepackage{setspace}
\usepackage{bigdelim}
\usepackage{bigstrut}
\usepackage{floatflt}	
\usepackage{wrapfig}
\usepackage{multicol}

\usepackage{indentfirst}
\usepackage[utf8]{inputenc}
\usepackage{tabularx}

\newcommand{\omegabar}{$\Omega_{\textrm{bar}}$~}
\newcommand{\omegaHI}{$\Omega_{\textrm{\h1}}$~}
\newcommand{\OLR}{$\Omega_{\textrm{\h1}} + \kappa/2$~}
\newcommand{\oUHR}{$\Omega_{\textrm{\h1}} + \kappa/4$~}
\newcommand{\ILR}{$\Omega_{\textrm{\h1}} - \kappa/2$~}
\newcommand{\iUHR}{$\Omega_{\textrm{\h1}} - \kappa/4$~}

\makeatletter
\newcommand*\mysize{%
   \@setfontsize\mysize{6.8}{10.0}%
}
\makeatother

\sffamily
\def\h1{H\,{\sc i}}
\def\hh{H$_2$}

\def\c1{C\,{\sc i}}

\def\NH3{NH$_{3}$}
\def\ch3cn{CH$_{3}$CN}

\def\deg{$^{o}$}

\def\kms{km s$^{-1}$}

\def\apjs{ApJSS}
\def\apjl{ApJL}

\def\aap{A\&A}

\defcitealias{obreschkow14}{OG14}
\defcitealias{obreschkow16}{O16}
\defcitealias{Murugeshan2019}{M19}
\defcitealias{Murugeshan20}{M20}

\title[\h1-RINGS]{The \h1 in Ring Galaxies Survey (\h1-RINGS) - Effects of the bar on the \h1 gas in ring galaxies}

\author{C. Murugeshan}
\affiliation{ATNF, CSIRO, Space and Astronomy, PO Box 1130, Bentley, WA 6102, Australia}
\email[C. Murugeshan]{chandrashekar.murugeshan@csiro.au}

\author{R. D\v{z}ud\v{z}ar}
\affiliation{Centre for Astrophysics and Supercomputing, Swinburne University of Technology, Hawthorn, Victoria 3122, Australia}
\alsoaffiliation{Tribalism PTY LTD, Lvl 1, 25 King Street Melbourne, 3000 Victoria, Australia} 

\author{R. Bagge}
\affiliation{School of Physics, The University of New South Wales, Kensington Campus, Old Main Building, Kensington, Sydney NSW Australia}
\alsoaffiliation{International Centre for Radio Astronomy Research, The University of Western Australia, 35 Stirling Highway, Crawley, WA 6009, Australia}

\author{T. O'Beirne}
\affiliation{International Centre for Radio Astronomy Research, The University of Western Australia, 35 Stirling Highway, Crawley, WA 6009, Australia}
\alsoaffiliation{ATNF, CSIRO, Space and Astronomy, PO Box 1130, Bentley, WA 6102, Australia}

\author{O. I. Wong}
\affiliation{ATNF, CSIRO, Space and Astronomy, PO Box 1130, Bentley, WA 6102, Australia}
\alsoaffiliation{International Centre for Radio Astronomy Research, The University of Western Australia, 35 Stirling Highway, Crawley, WA 6009, Australia}

\author{V. A. Kilborn}
\affiliation{Centre for Astrophysics and Supercomputing, Swinburne University of Technology, Hawthorn, Victoria 3122, Australia}
\alsoaffiliation{ARC Centre of Excellence for All Sky Astrophysics in 3 Dimensions (ASTRO 3D), Australia}

\author{M. E. Cluver}
\affiliation{Centre for Astrophysics and Supercomputing, Swinburne University of Technology, Hawthorn, Victoria 3122, Australia}
\alsoaffiliation{Department of Physics and Astronomy, University of the Western Cape, Robert Sobukwe Road, Bellville, 7535, South Africa}

\author{K. A. Lutz}
\affiliation{Observatoire Astronomique de Strasbourg, Université de Strasbourg, CNRS, UMR 7550, 67000 Strasbourg, France}

\author{A. Elagali}
\affiliation{Minderoo Foundation, 171 - 173 Mounts Bay Road, Perth, WA 6000, Australia}
\alsoaffiliation{School of Biological Sciences, The University of Western Australia, 35 Stirling Highway, Crawley, WA 6009, Australia}

\doi{10.1017/pasa.2020.32}

\received {dd Mmm YYYY}
\revised  {dd Mmm YYYY}
\accepted {dd Mmm YYYY}
\published{22 September 2020}


\begin{document}

\begin{abstract}
We present a new high-resolution neutral atomic hydrogen (\h1) survey of ring galaxies using the Australia Telescope Compact Array (ATCA). We target a sample of 24 ring galaxies from the Buta (1995) Southern Ring Galaxy Survey Catalogue in order to study the origin of resonance-, collisional- and interaction-driven ring galaxies. In this work, we present an overview of the sample and study their global and resolved \h1 properties. In addition, we also probe their star formation properties by measuring their star formation rates (SFR) and their resolved SFR surface density profiles. We find that a majority of the barred galaxies in our sample are \h1 deficient, alluding to the effects of the bar in driving their \h1 deficiency. Furthermore, for the secularly evolving barred ring galaxies in our sample, we apply Lindblad's resonance theory to predict the location of the resonance rings and find very good agreement between predictions and observations. We identify rings of \h1 gas and/or star formation co-located at one or the other major resonances. Lastly, we measure the bar pattern speed ($\Omega_{\textrm{bar}}$) for a sub-sample of our galaxies and find that the values range from 10 -- 90 \kms kpc$^{-1}$, in good agreement with previous studies.
\end{abstract}


\section{Introduction}

Galaxies come in a variety of shapes and sizes and have been traditionally classified on the basis of their optical morphology (e.g.,~\citealt{Hubble1926};~\citealt{de_Vaucouleurs1959}). Hubble developed the famous ``tunning-fork" diagram which sorted galaxies broadly into spheroidal/elliptical early-type galaxies towards one end of the fork, gradually transitioning into more and more late-type disky galaxies, which are observed to possess features such as a nucleus, bar and spiral arms to name a few. Galaxies were further classified on the basis of how tightly wound their spiral arms were and on the presence or absence of a bar, forming the two prongs in the tuning fork diagram. Apart from the early- and late-type galaxies, Hubble also identified galaxies that did not specifically fit into one or the other group and classified them as peculiar galaxies. In the following decades a number of atlases of galaxies were curated in an attempt to uniformly classify galaxies, one of the most important ones being the atlas compiled by \cite{Sandage1961}, who expanded on the existing Hubble classification scheme and introduced a special class of galaxies called ring galaxies. Such galaxies were observed to posses ring-like structures based on their optical images. This classification scheme was further revised by \cite{devaucouleurs91}, who introduced a number of sub-classes of ring galaxies, including those galaxies having nuclear, inner and outer (pseudo-) rings and lenses. 

Depending on the morphology of the ring, its position, alignment and size, ring galaxies are typically classified into the following major types: Resonance rings, collisional, accretion and polar rings (for a full review see~\citealt{Buta1996}). While the exact mechanisms driving the formation of rings in galaxies are still debated, the following are believed to be some of the major scenarios: 

\textbf{1. Resonance rings driven by the bar:} Stars and gas orbiting the inner parts of a galaxy possessing a bar respond to the torques exerted by the non-axisymmetric potential of the bar and settle into orbits of resonance (\citealt{Lindblad1964};~\citealt{Schwarz1981}). In addition to the accumulation of stars and gas from the disk, due to the increased density and pressure at these resonance locations, the gas collapses to form new stars, resulting in the formation of stellar ring-like structures that is visible in the UV, optical and infrared (IR) bands (\citealt{Buta1996};~\citealt{Buta2004};~\citealt{Grouchy2010} and references therein). Numerous sub-classes exist within resonance rings, such as nuclear rings, inner (pseudo) rings and outer (pseudo) rings (e.g.~\citealt{Comeron2014}). Outer pseudo rings are observed to be formed by two tightly wound spiral arms in late-type galaxies that almost form a complete ring in the outskirts of galaxies \citep{Buta1993}. Pseudo rings are purported to be reflective of an earlier stage in ring formation in galaxies, as they are more common in late-type galaxies, while complete rings are more common among early-type disk galaxies (\citealt{Elmegreen1992};~\citealt{Buta1993};~\citealt{Buta1996}). The theoretical models of \cite{Schwarz1981} however predict both rings and pseudo rings to occur close to the Lindblad resonance radii in galaxies. We present a more detailed discussion of resonance rings in Section~\ref{sec:Bar_driven_resonance_rings}. \\

\noindent \textbf{2. Driven by galaxy-galaxy collisions:} Simulations have shown that galaxy collisions could lead to the formation of ring galaxies (\citealt{Appleton96};~\citealt{Wong2006};~\citealt{Elagali2018_eagle} and references therein). In a drop-through collision, typically a smaller ``bullet'' galaxy passes through the (off-)centre of a larger disk galaxy, sending out shock waves that push both the stellar and gas material outward to form a ring-like structure. Many such systems have thus far been observed including the famous Cartwheel galaxy AM 2026-424 \citep{Zwicky41}, and the recently discovered high-redshift collisional ring galaxy R5519 \citep{Yuan20}. \\

\noindent \textbf{3. Driven by accretion of gas from the IGM:} In few ring systems such as Hoag’s galaxy \citep{Hoag50}, the Leo ring \citep{Schneider83}, AGC 203001 \citep{Bait20}, the formation of the ring structure is attributed to the accretion of fresh gas from the surrounding IGM onto a ring, under favourable conditions \citep{Lynds76, Theys77, Schweizer87}. Star formation starts as soon as the accreted \h1 gas column density exceeds a threshold. In some cases, for example the Leo ring and AGC 203001 \citep{Bait20}, the accreted \h1 gas column density is low enough that significant star formation does not occur, leaving only an \h1 ring behind with no stellar counterpart. Polar rings are another class of ring galaxies which are believed to have accreted \h1 gas, but that gas has settled into polar orbits around the galaxy. Typically, polar ring galaxies are observed to be S0-type galaxies which have recently accreted fresh \h1 gas from either a smaller companion dwarf galaxy or directly from the IGM. To date over 400 potential polar ring galaxy candidates have been identified, but only a dozen or more galaxies have been confirmed to be polar rings (\citealt{Moiseev2011};~\citealt{Egorov2019};~\citealt{Nishimura2022}). As such, polar ring galaxies are a rare class of ring galaxies. \\

Ring galaxies have been studied extensively in the optical, UV, NIR and FIR bands. Early photometric studies revealed that the location of the rings were hot-spots for active star formation, which was determined from their distinct patchy blue distribution (\citealt{Sandage1961};~\citealt{Athanassoula1985}).
~\cite{Buta1996} find that among the catalogue of bright nearby galaxies, outer (pseudo)rings are found only for about 10\% of the sample. \citet{Buta1995} compiled a comprehensive list of 3692 ring galaxies, with the objective of studying if the rings in galaxies were linked to orbital resonances. They speculate that the observed outer (pseudo)rings may be associated with the outer Lindblad resonance, while the inner rings are associated with the Ultraharmonic resonances. They suggest that measuring the location of the inner and outer rings may be useful to indirectly measure the bar pattern speed. For a full review of the various studies undertaken in the different wavelength regimes we refer the reader to \cite{Buta1996}.

In terms of the \h1 gas in ring galaxies, most studies have focused on the unresolved, global \h1 gas properties with a handful of high-resolution \h1 studies limited to small samples. One of the first high-resolution study of the \h1 gas distribution in a ring galaxy was undertaken by \cite{Bosma1977} for the spiral galaxy NGC 4736, via synthesis imaging using the Westerbork Radio Synthesis Telescope (WSRT). They find that the \h1 gas in this galaxy is observed to trace not only the optically bright central regions, but also the faint outer stellar ring. Following this, a few targeted \h1 studies focusing on a few galaxies were conducted, most of which revealed \h1 rings to be roughly co-located with the outer stellar rings (see for example \citealt{Knapp1984};~\citealt{van_Driel1988};~\citealt{van_Driel1991}). 
Some early \h1 studies have also identified galaxies for which rings of \h1 gas are found with no obvious optical ring-like structures (see for example \citealt{Krumm1985};~\citealt{Tilanus1993}). Overall, such studies have hinted at the importance of the \h1 gas in forming the ring-like stellar structures observed in galaxies.

Apart from a few targeted \h1 studies, to date, there is no large comprehensive study of the resolved \h1 gas distribution and kinematics of ring galaxies. \h1 gas is an excellent tracer of both internal and external processes affecting galactic evolution. In addition, \h1 gas is the fundamental source of molecular hydrogen (\hh), which is eventually converted to stars in low-z galaxies \citep{Meurer2006}. This link proves even more useful in the context of tracing star formation in the different types of ring galaxies. Therefore, such a study will be an important step towards understanding how the \h1 gas in these systems is distributed and also enable us to study the effects of the environment on such systems. This may be particularly useful in the context of understanding ring formation in galaxies linked to tidal interactions and external perturbations (e.g.,~\citealt{Combes1990}). In addition, from the gas kinematics we can measure the angular momentum (AM) in the ring and examine the link between gas stability and the role of AM in regulating star formation in the ring. \h1 studies of ring galaxies are therefore very important to understand both the morphology and kinematics of the rings, allowing us to understand the various processes (both internal and environmental) driving their formation.

It is important to distinguish mechanisms that lead to the formation of ring galaxies (visible in their optical and \h1 morphology) as discussed above compared to \h1~``holes" and \h1~``depressions" at the centers of galaxies that may be linked to the natural process of star formation where the \h1 gas is converted to \hh~ and eventually stars (see~\citealt{Walter2008};~\citealt{Leroy2008};~\citealt{Murugeshan2019}) and do not necessarily coincide with the optical rings. In this paper, we introduce a sample of ring galaxies which have been identified through their optical morphology to possess ring-like structures (see Section~\ref{subsec:Sample_selection} for more details on the selection criteria). We probe the distribution of the \h1 gas in such systems and study their kinematics. In the next sub-section we introduce this new survey of ring galaxies.

\subsection{The HI in Ring Galaxies Survey (HI-RINGS)}

In order to address the diversity of ring galaxy evolution in terms of the gas kinematics, we have put together a new high-resolution \h1 survey of ring galaxies -- the \h1 in Ring Galaxies Survey (\h1-RINGS),  targeting different types of ring galaxies, to study their \h1 and star formation properties using a uniform and large sample of 24 galaxies. We have procured new Australia Telescope Compact Array (ATCA) observations for 15 ring galaxies in addition to compiling ATCA data from the archive for 9 galaxies. Our sample consists of various types of ring galaxies including resonance, collisional and interaction rings, enabling us to probe their \h1 gas distribution, resolved star formation and star formation rate (SFR) surface densities, as well as their kinematic properties, using a multiwavelength approach including optical and infrared data. 

Through this paper and the following series of papers, we hope to address important questions pertaining to the formation and evolution of ring galaxies in the local Universe. Below we list some of the main goals of the \h1-RINGS survey.
\begin{itemize}
   
     \item Test Lindblad's resonance theory using robust rotation curves for a large sample of ring galaxies. We will utilise the rotation curves to derive the bar pattern speeds and predict the location of the resonance rings. In addition, we aim to study the \h1 gas and SFR surface densities to probe the nature of the distribution of both gas and star formation in our sample galaxies.
    
    
    \item We aim explore the \h1/H$_2$ ratio in ring galaxies. It has been observed in several collisional ring galaxies that the \h1/H$_2$ ratio is large, this could mean that in such galaxies the H$_2$ is being efficiently converted into stars (\citealt{Higdon11}; \citealt{Wong17}). We intend to compare the \h1/H$_2$ ratio among the different types of ring galaxies in our sample to understand the \h1 to H$_2$ conversion phase in such systems. More than half the galaxies in our sample have high-resolution ALMA observations of CO. By complementing this with high-resolution \h1 data, we will be able to study the \h1/H$_2$ ratios for a large sample of spatially resolved ring galaxies.
    
    \item \citet{Obreschkow2016} introduced the $f_{\mathrm{atm}} - q$ relation, where $f_{\mathrm{atm}}$ is the neutral atomic gas fraction and $q \propto j_{\mathrm{b}}/M_{\mathrm{b}}$ is the integrated disk stability parameter, where $M_{\mathrm{b}}$ is the total baryonic mass  and $j_{\mathrm{b}}$ is the total specific baryonic angular momentum (sAM) of the galaxy. Both \h1-excess and -deficient late-type galaxies have been found to follow this relation consistently \citep{Lutz17, Lutz18, Murugeshan2019}, indicating that sAM plays a fundamental role in regulating the \h1 gas content in galaxies. \citet{Murugeshan2019} find two ring galaxies in their sample that follow the $f_{\mathrm{atm}} - q$ relation. However, do all galaxies possessing rings that are in gravitational equilibrium follow the $f_{\mathrm{atm}} - q$ relation? We will use the high-resolution \h1 data to accurately model the kinematics of ring galaxies, measure their sAM and study their behaviour on the $f_{\mathrm{atm}} - q$ plane. Such a study will give us important clues about the role of angular momentum in regulating the stability of the \h1 gas in the ring.
    
    \item What is the effect of the local and global environment on the formation of ring galaxies? \citet{Elagali2018_eagle} found that 83\% of ring galaxies in the EAGLE simulations were formed via a `drop-through' collision. 
    Our sample of ring galaxies spans across a range of local density environments, being associated with the field, group or cluster. With our sample of ring galaxies, we aim to study how the environment affects the properties of ring galaxies. 
    
\end{itemize}

In this work, we introduce the \h1-RINGS survey and give an overview of the sample. We model the \h1 kinematics for the sample galaxies using 3D tilted-ring fitting and derive robust rotation curves. Using the rotation curves we accurately derive the bar pattern speed for the barred galaxies in our sample and test Lindblad's theory of ring formation. In addition, we also explore the effects of the bar on the resolved and global \h1 and star formation properties of galaxies. We structure the paper as follows -- in Section~\ref{sec:Data&Methods} we give details on the sample selection, observations, data reduction, source finding and 3D modelling. Following this, in Section~\ref{sec:Results} we present the main results, including an overview of the sample and their measured properties. In Section~\ref{sec:Bar_driven_resonance_rings}, we test resonance theory of ring formation based on the analysis of a sub-sample of barred galaxies in our sample. Finally, in Section~\ref{sec:Summary_Future}, we summarise the main results and give an outline of the plans for the future.

\section{Data and Methods}
\label{sec:Data&Methods}

In this section we will discuss the criteria applied for the sample selection, give details of the observations, list the steps followed in the data reduction process and discuss the source finding and characterisation. We will also describe the kinematic modelling for the sample galaxies which is necessary to derive their rotation curves.

\subsection{Sample selection}
\label{subsec:Sample_selection}
We use the Catalogue of Southern Ringed Galaxies (CSRG, \citealt{vizierButa}), obtained through the Vizier catalogue access tool \citep{vizier} to select our sample of ring galaxies. The CSRG was originally published by \citet{Buta1995}, presenting properties of outer and inner rings, pseudo-rings and lenses of 3692 galaxies south of -17$^\circ$ in declination.

Our selection criteria was based on galaxy size, position, visual inspection of each galaxy from CSRG, and a cross-check as to whether the galaxies were detected in the \h1 All Sky Survey (HIPASS,~\citealt{Barnes2001};~\citealt{Meyer2004}):
\begin{itemize}
    \item Our primary selection criteria was galaxy size, selecting galaxies with large angular size, sufficient to resolve the \h1\ in the rings, to be able create reliable kinematic models. We selected galaxies with D$_{\textrm{25}}$ $\geq$ 180$^{\prime\prime}$, where D$_{\textrm{25}}$ is the B-band 25 mag~arcsec$^{-2}$ isophotal diameter, from ESO-LV "Quick Blue" IIa-O \citep{ESOQuick}.
    
    \item The secondary constraint was declination of the galaxy, selecting galaxies south of -25$^\circ$ in declination. This criteria ensures good observability with the ATCA, avoiding an elongated synthesised beam.
    
    \item Using Astroquery SkyView cutout service \citep{Astroquery19}, we obtained DSS images of all $\sim$3692 galaxies from the CSRG and visually inspected them. This step was necessary to validate the galaxy selection process and it was done before and after size/declination restrictions to the catalogue. Next, with visual inspection of 101 ring galaxies -- obtained after the declination and size restrictions -- we removed highly edge-on galaxies and galaxies with no clear rings and/or inner (pseudo)rings. Galaxies with unclear/faint images were re-checked in NED database. Additionally, two ring galaxies -- ESO 215-31 and ESO 179-IG-013, were added to our sample through visual inspection. 
    
    \item As a final check, we inspected the HIPASS data to see if the selected galaxies have an \h1\ detection. This step was necessary to minimize risk of not detecting \h1\ through our new ATCA observations. With this check five galaxies were removed due to \h1 non detection in HIPASS. 
\end{itemize}

\begin{figure}
    \centering
    \includegraphics[width=1\columnwidth]{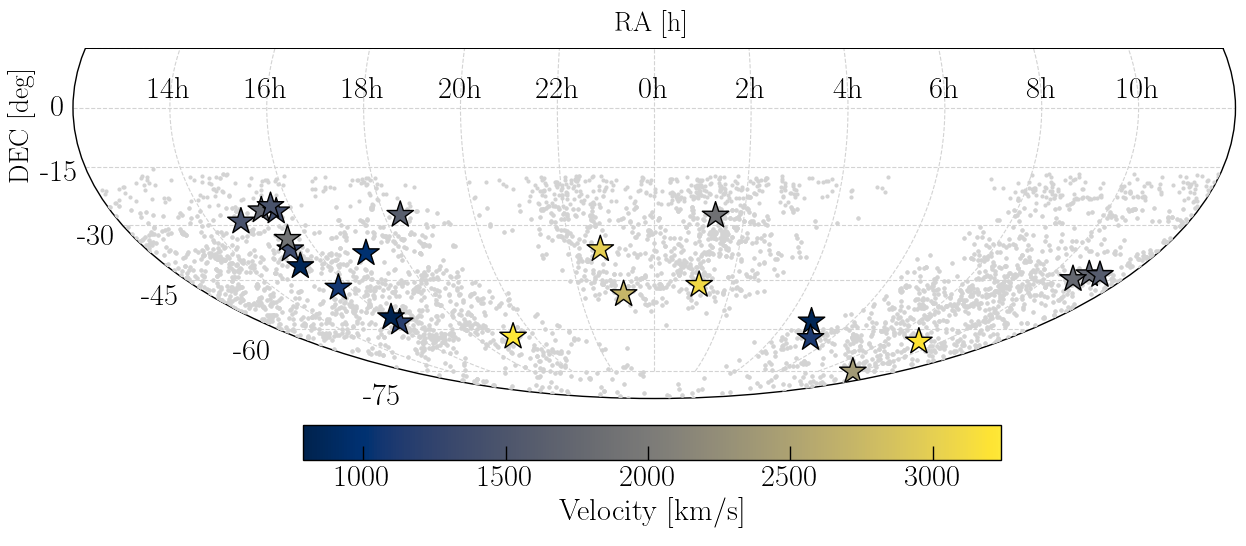}     \caption{Spatial and velocity distribution of galaxies in the \h1 in Ring Galaxies Survey (\h1-RINGS). The grey points show all the ring galaxies from the \citep{Buta1995} catalogue, the stars show our sample of ring galaxies where colour denotes their Heliocentric velocity.}
    \label{fig:SelectedRingGalaxies}
\end{figure}

The above criteria left us with a sample of 27 ring galaxies of the $\sim$3692 from the original CSRG sample, as the majority of galaxies have an angular size too small to be sufficiently resolved with the ATCA. Out of the 27, 12 galaxies were found to have archival ATCA data (in the 1.5km, 750m and 352/375m configurations). We have observed 15 ring galaxies with our ATCA proposal (C3385). Ultimately, due to calibration issues and/or not being able to derive reliable kinematic models, 3 of the 12 galaxies from the archive were rejected, leaving us with a final sample size of 24 ring galaxies. Fig.~\ref{fig:SelectedRingGalaxies} shows the spatial and velocity distribution of our sample of ring galaxies.

\subsection{Observations}
\label{subsec:observations}
The 15 galaxies in our sample, for which no archival data are available, were observed with the ATCA in one of the default 750m and 1.5km array configurations. These observations were carried out through the ATCA project C3385 in 2020OCTS and 2021APRS semesters with a total observation time of 307.5h, and an average on-source time of $\sim 11$h per array configuration. Combining the data from these two configurations will result in increasing both sensitivity and spatial resolution of the synthesised data, which is key to accurately modelling the kinematics of the sample galaxies (see Section~\ref{subsec:3D_modelling}). The ATCA, which is equipped with the Compact Array Broad-band Backend (CABB, \citealt{Wilson2011}), allows us to observe the galaxies in both continuum and spectral-line mode. The total bandwidth in the continuum band is 2048 MHz with a frequency resolution of 1 MHz, centered at 2.1 GHz. The spectral line observations have a total bandwidth of 8.5 MHz with a channel resolution of $0.5$ kHz, which corresponds to a velocity resolution of $\sim 1$ \kms. We employed the following strategy for the observations -- 55 min on source followed by 5 min on the phase calibrator each hour. The primary calibrators used are PKS 1934-638 and PKS 0823-500 depending on the time of observations. The primary calibrator was observed for 10 min at the start and end of the observing run. Other details of the observations for the sample are listed in Table~\ref{tab:Obervations}.

\subsection{Archival ATCA data}
\label{subsec:archival_data}

As mentioned previously, 9 galaxies in our sample have archival ATCA data in both the standard 1.5km and 750m array configurations. The data was retrieved from the Australia Telescope Online Archive (ATOA). For a few galaxies we also found observations in the 350m array configuration. The exact configurations used for the data reduction for the various galaxies in our sample is listed in Table~\ref{tab:Obervations}.    

\begin{table*}
\caption{Details of the observations. Col (1): Common name; (2) ATCA baseline observation configuration; Col (3): The central frequency of the observation; Col (4): Primary flux calibrator used; Col (5): Phase calibrator used; Col (6): rms in the image cubes; Col (7): Synthesized beam size major and minor axis, respectively; Col (8): Notes about observations: This survey - ATCA project C3385; NGC 1433 data is from C305 \citep{Ryder1996}; NGC 7531 data is from C2196 \citep{Koribalski2009}; NGC 1808 data is from C585 \citep{Dahlem2001}; NGC1533 data is from C934 and C1003 \citep{Ryan-Weber2003}; NGC 7098 data is from C744 \citep{Pisano2002}}

\begin{tabular}{llclllcl}
\hline\hline
Name  & Array & $\nu_{0}$ &  Primary   & Secondary & rms & Synthesized beam  & Notes \\
     & Configurations & [MHz]  & Calibrator  & Calibrator & [mJy beam$^{-1}$] & [arcsec] &  \\
(1)  & (2)  & (3) &  (4)  & (5) & (6)  & (7) & (8) \\
\hline\hline
ESO 215-31 &  750m \& 1.5km & 1407 & 1934-638 & 0823-500 & 0.67 & 43$\times$30  & This survey \\

ESO 269-57 &  750m \& 1.5km & 1405 & 1934-638 \& 0823-500 & 1320-446 & 2.20  & 43$\times$30 & This survey \\

NGC 1326 & 750m \& 1.5km & 1413 & 1934-638 & 0332-403 & 1.06 & 93$\times$25 & This survey \\

NGC 3358 &  750m \& 1.5km & 1406.3 & 1934-638 & 1015-314 & 1.85 & 54$\times$31 & This survey \\

IC 5267 & 750m \& 1.5km & 1412.2 & 1934-638 & 2311-452 & 1.09 & 49$\times$29 & This survey \\

NGC 1302 & 750m \& 1.5km & 1412 & 1934-638 & 0237-233 & 1.53 & 73$\times$28  & This survey \\

NGC 1398 & 750m \& 1.5km & 1413 & 1934-638 & 0413-210 & 1.26 & 68$\times$27 &  This survey\\

NGC 2217 & 750m \& 1.5km & 1412 & 1934-638 & 0614-349 & 2.13 & 74$\times$30 & This survey\\

NGC 7020 & 750m \& 1.5km  & 1405.2 & 1934-638 & 2117-642 & 1.67 & 38$\times$30 & This survey\\

NGC 1291 &  750m \& 1.5km & 1416 & 1934-638 & 0332-403 & 1.62 & 48$\times$29 & This survey\\

NGC 1371 & 1.5km & 1413.2 & 1934-638 & 0413-210 & 2.58 & 50$\times$19 & This survey\\

ESO 179-IG-013 & 750m \& 1.5km & 1416 & 1934-638 & 1613-586 & 2.09 & 36$\times$29 & This survey\\

NGC 1079 & 1.5km & 1413.2 & 1934-638 & 0237-233 & 2.41 & 57$\times$19 & This survey\\

NGC 5101 & 1.5km & 1411.7 & 1934-638 & 1308-220 & 2.71 & 54$\times$19 & This survey\\

IC 5240 &  750m \& 1.5km & 1411.7 & 1934-638 & 2226-411 & 0.99 & 48$\times$28 & This survey\\

NGC 1350 & 750m \& 1.5km & 1415.5 & 1934-638 & 0237-233  & 0.94 & 57$\times$23 & \cite{Murugeshan2019} \\

NGC 2369 & 750m \& 1.5km & 1409 & 0823-500 & 0637-752 & 2.33 & 31$\times$24 & \cite{Murugeshan2019} \\

NGC 7098 & 375m \& 750m & 1410 & 1934-638  & 2142-758 & 4.09 & 74$\times$58 & C744 \\

NGC 1543 & 750m \& 1.5km & 1419 & 1934-638 & 0420-625 & 1.25 & 37$\times$29 & \cite{Murugeshan2019} \\

NGC 1533 & 375m, 750m \& 1.5km & 1417 & 1934-638 & 0407-658 & 8.2 & 35$\times$27 & C934 and C1003 \\

NGC 1808 & 750m \& 1.5km & 1414 & 1934-638 & 0521-365 & 1.00 & 50$\times$30 & C585 \\

NGC 7531 & 352m \& 1.5km & 1412.8 & 1934-638 & 2311-452 & 3.29 & 91$\times$62 & C2196 \\

NGC 1433 & 750m \& 1.5km & 1415 & 0438-436 & 0438-436  & 1.32 & 46$\times$34 & C305 \\

NGC 6300 & 750m \& 1.5km & 1415 & 1934-638 & 1718-649 & 0.96 & 36$\times$22 & \cite{Murugeshan2019} \\

\hline
\end{tabular}
\label{tab:Obervations}
\end{table*}

\subsection{Data Reduction}
\label{subsec:Data_reduction}

We used the \texttt{MIRIAD} software package \citep{Sault1995} to reduce the ATCA data. The 750m and 1.5km array data (and for some galaxies the 375m and 352m data) were reduced separately and later combined to produce the final visibility data set. The following steps are carried out in the data reduction process -- first we flag any spurious data using \texttt{UVFLAG} and \texttt{PGFLAG}. Once the quality of the data looks good, we then perform the bandpass and phase calibration using the tasks \texttt{MFCAL}, \texttt{MFBOOT}, \texttt{GPCAL} and \texttt{GPBOOT}. This is then followed by continuum subtraction using \texttt{UVLIN}. We iteratively select line-free channels in the data for the continuum subtraction process. 
Following this, we use the \texttt{INVERT} task to fourier transform the UV data to form a dirty image. The visibility data from the 1.5km and 750m configurations are combined using the \texttt{INVERT} task after continuum subtraction has been carried-out on the individual data sets. The \texttt{INVERT} task also estimates the expected theoretical rms noise in the image cube. We use the 3$\sigma$ value of this estimate as cutoff during the ``cleaning'' process. Next, we proceed to the de-convolution step, which is performed using the \texttt{CLEAN} algorithm in \texttt{MIRIAD}. We set the number of iterations for the minor \texttt{CLEAN} cycle to 2000 and wait for \texttt{CLEAN} to reach the cutoff value. Following the de-convolution step, we restore the final image cube using the \texttt{RESTOR} task. Finally, we apply the primary beam correction to the restored image cube using the \texttt{LINMOS} task.

We set the robustness parameter to 0.5, as it offers a good compromise between sensitivity and resolution. The semi-major axis values of the synthesized beam (BMAJ) range from 31 to 93 arcsec, with a median beam size of $\sim 40$ arcsec. The channels were also smoothed to a width of 5 \kms, thereby increasing the signal-to-noise-ratio (SNR) in the image cubes. The median 3$\sigma$ rms noise level in the image cubes for our galaxy sample is $\sim 4.92$ mJy/beam per channel, which corresponds to a column density sensitivity limit of N$_{\textrm{\h1}} \sim 1.7 \times$10$^{19}$cm$^{-2}$ assuming an observed median beam size of $\sim 40$ arcsec.

\subsection{Source Finding with SoFiA}
\label{subsec:Source_finding}

We make use of the Source Finding Application (SoFiA;~\citealt{Serra15};~\citealt{Westmeier21}) to automatically detect sources in the final image cubes. We have used the smooth + clip (S + C) algorithm in SoFiA to build the source mask. The S + C finder works by spatially and spectrally smoothing the data cube on multiple scales as defined by the user. In each smoothing iteration, a user-specified flux threshold relative to the global rms noise level after smoothing is applied to the data, and all pixels with an absolute flux density exceeding that threshold is used to create the source mask \citep{Westmeier21}. The multi-scale smoothing and clipping helps in identifying relatively faint and extended features in the data cubes that may have been missed otherwise. After the source finding run, SoFiA returns a number of data products including a catalogue with source-specific parameters such as their RA, Dec, systemic velocity (V$_{\textrm{sys}}$), integrated flux (S$_{\textrm{int}}$), $w_{20}$ and $w_{50}$ velocity widths and kinematic position angle (P.A) based on the \h1 gas kinematics. In addition, SoFiA also produces cutouts of the spectral cube, moment 0 (intensity), moment 1 (velocity) and moment 2 (dispersion) maps for all detected sources from the inputed image cube. We make use of the spectral cubes and moment maps derived from SoFiA for the rest of our analysis.

\subsection{Interactive visualisation of the HI-RINGS galaxies}
Using Streamlit, an open-source Python library, we have built a web application for interactive visualization of our \h1-RINGS survey\footnote{Link to the interactive web app: \url{https://hi-rings.streamlitapp.com/}}. The web application allows the user to interactively explore our sample of ring galaxies using \h1 intensity data together with data from several other surveys including the DSS, WISE, 2MASS and GALEX. Within the app, it is possible to overlay selected \h1 contours on top of image cutouts from the different surveys. The app also allows image comparison with and without \h1 contours, and image comparison with different survey images in the background. We believe such an interactive tool will be both informative as well as provide important visual context into the distribution of the \h1 and stellar components of the different types of ring galaxies in our sample. We provide further details on the web application in \ref{appendix:streamlit_app}. In addition, we also intend to host the reduced \h1 data-products for the \h1-RINGS sample.

\subsection{Tilted Ring Modelling with 3DBarolo}
\label{subsec:3D_modelling}

We perform 3D tilted-ring fitting \citep{rogstad74} to the \h1 emission observed in the data cubes for our sample galaxies using the publicly available software \texttt{3DBarolo}~\citep{DiTeodoro2015}. \texttt{3DBarolo} takes in a user-specified parameter file containng initial guesses for the fitting and outputs the best fit tilted-ring model for the galaxy. For the fitting strategy we follow the steps described in \citet{Murugeshan2019}. \texttt{3DBarolo} fits $n$ independent tilted rings (where $n$ is the total number of resolution elements across the major axis) of width equivalent to half the FWHM of the resolution element (which varies from galaxy to galaxy in our sample). The fitting is performed in two stages -- in the first stage the user-defined initial guesses for the galaxy centre (set to the optical/NIR centres obtained from NED\footnote{\url{http://ned.ipac.caltech.edu/}}), the systemic velocity (V$_{\textrm{sys}}$) and position angle (PA) are set to the values determined from the SoFiA source finding run. The initial guess for the inclination angle ($i$) is set to the value determined by \citet{Lauberts89} based on the $B$-band optical surface photometry of the galaxies. We allow PA, $i$ and the rotation velocity ($v_\textrm{rot}$) to vary as free parameters during the fit. Once the best fitting model is derived from the first stage, the code uses these values in the second stage of the fitting. Finally, the best fitting values for PA, $i$ and $v_\textrm{rot}$ for every ring is written-out to a log file. 
The best fit model parameters such as $v_{\textrm{rot}}$, $i$ and PA are shown for the full sample in Fig.~\ref{fig:parameter_plots1} and Fig.~\ref{fig:parameter_plots2} in ~\ref{appendix:notes_on_galaxies}.

We do note that since we are dealing with ring galaxies and deriving their rotation curves from resolved \h1 observations, there is the danger of missing the \h1 signal from the very centres of these galaxies (assuming the \h1 gas is primarily co-located with the optical ring). Upon investigation, we find that this is not the case for the majority of our sample galaxies. We find that in most cases there is still \h1 emission from their centres and that the signal-to-noise ratio (SNR) of the \h1 emission is high enough for 3DBarolo to accurately fit the rotation curve in the inner regions. We find four galaxies in our sample, namely, NGC 1326, NGC 1371, NGC 5101 and NGC 1534, for which the SNR of the \h1 emission at the centre is too low for 3DBarolo to effectively create a source mask and fit a kinematic model. We note that this might potentially affect the resonance analysis we perform in Section~\ref{sec:Bar_driven_resonance_rings} for the above galaxies. This is because the prediction of the resonance locations are dependent on their rotation curves. That being said, since the location of the resonance rings are mainly dependent on the outer parts of the rotation curve, we believe this should not significantly affect our analysis.

\section{Results}
\label{sec:Results}

We now discuss some general properties of the ring galaxies in our sample as well as present the main results.

\subsection{Flux comparison with HIPASS}
\label{subsec:flux_compare}

\begin{figure}
\hspace{-0.5cm}
    \includegraphics[width=9cm,height=7cm]{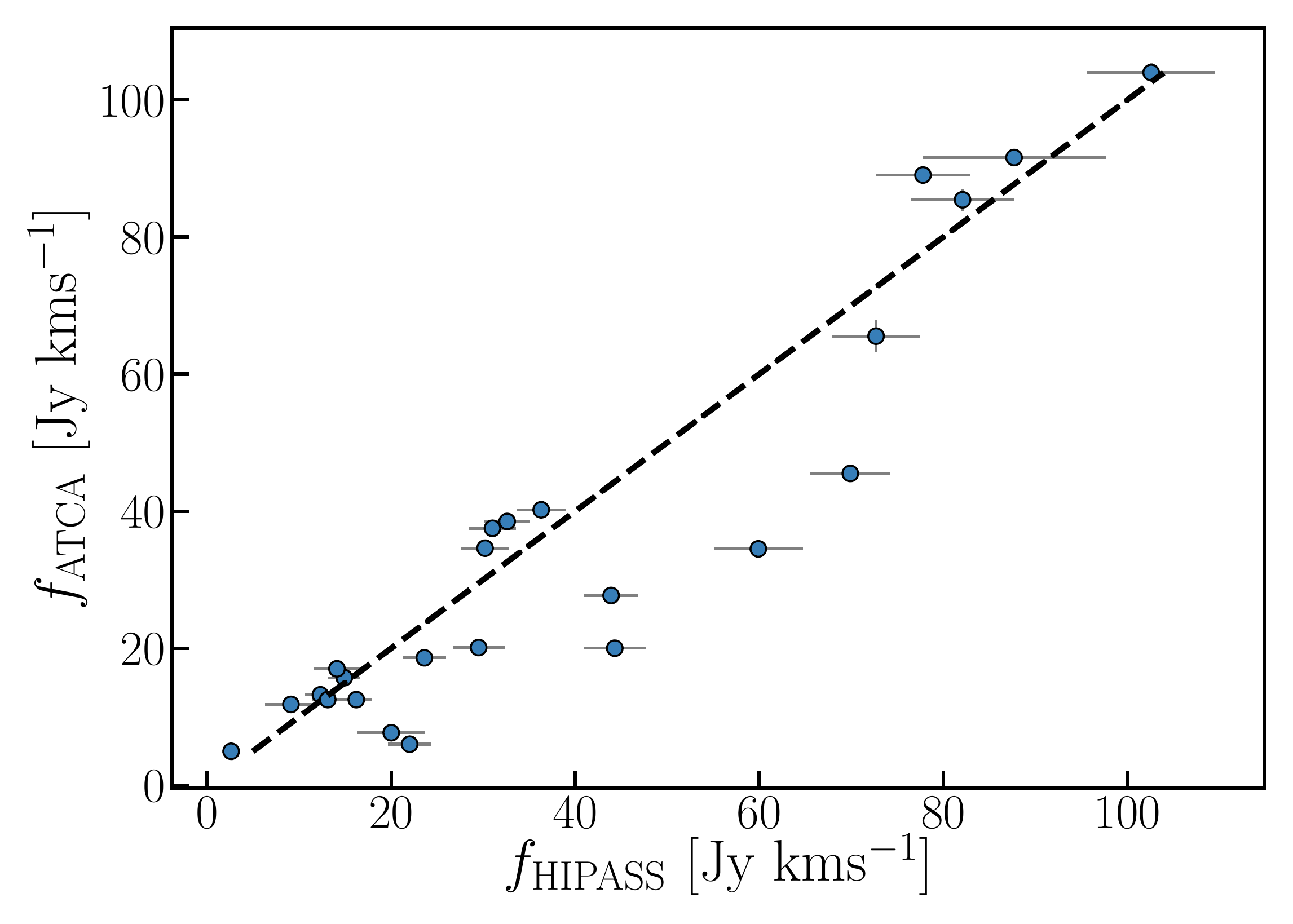}
    \caption{The plot compares the integrated flux for the ring sample derived from our ATCA observation (on the Y-axis) against their corresponding integrated flux from HIPASS (X-axis). The dashed black line represents the one-to-one correspondence.}
    \label{fig:flux_compare}
\end{figure}

As described in Section~\ref{subsec:Source_finding}, we use SoFiA to automatically detect the sources in the image cubes and also obtain a catalogue consisting of relevant \h1 parameters including the integrated flux. We check the quality of the data and as a validation step compare the integrated flux derived from the SoFiA run for our sample galaxies with their HIPASS flux values. Fig.~\ref{fig:flux_compare} shows a plot comparing the integrated flux from the ATCA observations with their corresponding HIPASS fluxes. For the majority of the sources, we find a good one-to-one correspondence. About 7 galaxies in our sample ($\sim 29\%$) are observed to have a low integrated flux compared to their HIPASS flux. We attribute this to a combination of two effects: 1) Missing diffuse flux -- Three galaxies NGC 1371, NGC 1079 and NGC 5101 only have observations in the 1.5km array configuration. This means that these galaxies are very likely missing the extended diffuse emission that the shorter baselines may have otherwise captured. This is also evidenced by the fact that the HIPASS $w_{20}$ values are typically $\sim 9\%$ broader than the ATCA $w_{20}$ values; 2) Confusion in the HIPASS flux -- Galaxies NGC 1326 and NGC 3358 are observed to have close companions spatially and/or in velocity space. Owing to the larger $\sim 15$ arcmin HIPASS beam, it is likely that the HIPASS flux is confused and overestimated for these two galaxies. This is also clearly seen in Fig.~\ref{fig:spectra_1}, where we compare their HIPASS and ATCA spectra. We examine the rms noise level in the image cubes and find them to be between 0.67 -- 8.2 mJy/beam per channel. This is close to the expected theoretical noise given the integration time and UV-coverage of the individual observations. In the following sections we present the main results from our study.

\subsection{Sample Overview and HI properties}
\label{subsec:Sample_overview}

In this section we give an overview of both the global and resolved properties of the ring galaxies in our sample. Figs.~\ref{fig:mosaic1} -- \ref{fig:mosaic4} in \ref{appendix:notes_on_galaxies} show the optical images of the sample galaxies with the \h1 contours overlaid on top, as well as their \h1 velocity maps. From the optical images, we find that most of the galaxies in our sample are early late-type galaxies with their morphology ranging from (SA, SAB to S0). In addition to the presence of optical rings, we also notice the \h1 gas distributed in the form of rings, in many cases (almost) co-located with the optical rings (NGC 1326, NGC 1398, NGC 2217, IC 5240, NGC 1350, NGC 6300, ESO 215-31, NGC 1079,  NGC 5101). We find this to be true typically for galaxies possessing bars in our sample. However, it is also interesting to note that the \h1 gas distribution in some ring galaxies is more complex. For example, galaxies ESO 215-31 and NGC 1808 are observed to have an excess of \h1 gas clumping at the ends of the bar and along the major axis of the bar, which may be indicative of gas inflows (see for example \citealt{Salak2016} and references therein). From the velocity maps it is evident that most galaxies in our sample exhibit ordered rotational motion.

We make use of the Carnegie-Irvine Galaxy Survey (CGS;~\citealt{Ho2011}) and the Third Reference Catalogue of Bright Galaxies \citep{devaucouleurs91} to confirm presence or absence of a bar for our sample galaxies. We visually observe that 14 galaxies in our sample have a strong bar, while the other 10 are observed to have a weak or no bar. For our analysis in the next sections, we therefore split the galaxies into two groups -- those with a strong bar and those with a weak and/or no bar. The \h1 mass for our sample galaxies is measured using the relation described in \citet{Walter2008}, \\

$M_{\mathrm{HI}}\left[\mathrm{M}_{\odot}\right]=2.356 \times 10^5 D^2 S_{\text {int }}$, \\

where $D$ is the distance in Mpc and $S_{\text {int }}$ is the ATCA integrated flux in units of Jy~\kms. The uncertainties on $M_{\mathrm{HI}}$ are computed by propagating the errors associated with $S_{\text {int }}$ as described in \citet{Koribalski2004} and assuming a 10\% uncertainty on the distance measures. The \h1 mass of the sample galaxies is in the range $8.61 \leq \log(M_{\textrm{\h1}}/M_{\odot}) \leq 10.19$, with a median \h1 mass of
 $\log(M_{\textrm{\h1}}/M_{\odot})$ $\sim 9.36$. 
 
We estimate the stellar mass for our sample using their 2MASS \citep{Skrutskie2006} $K_s$-band magnitudes as described by Eq. 3 in \citet{Wen2013}. The derived stellar mass for our sample is in the range $9.77 \leq \log(M_{\star}/M_{\odot}) \leq 11.07$, with a median value of $\log(M_{\star}/M_{\odot}) \sim 10.56$. In Table~\ref{tab:sample-properties} we list all the relevant derived quantities such as the integrated flux, \h1 and stellar mass etc, for our sample.

\begin{figure*}
\hspace{-0.8cm}
\begin{subfigure}{.5\textwidth}
  \centering
  \includegraphics[width=9cm,height=7cm]{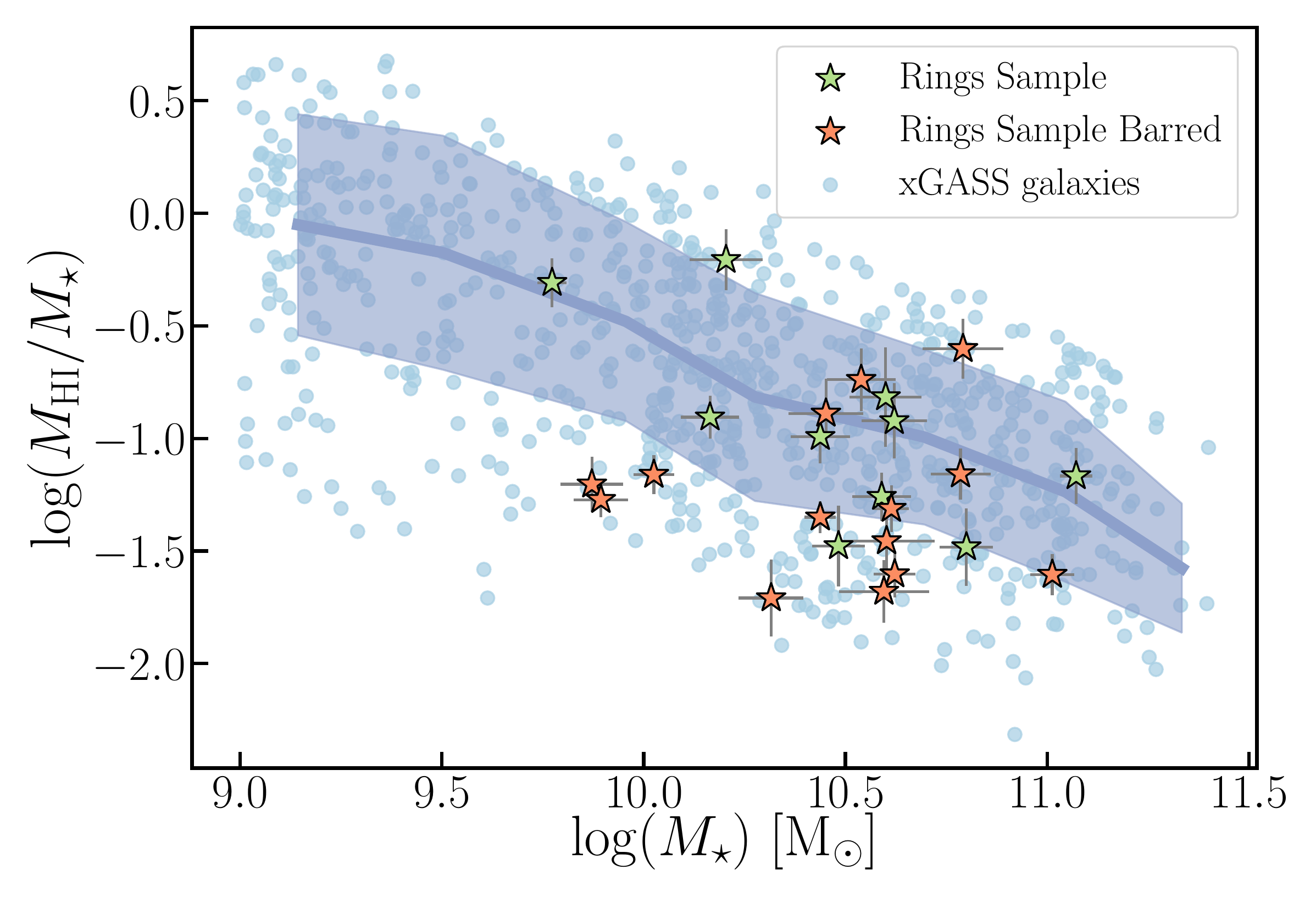}
  \caption{}
  \label{fig:gasfraction_a}
\end{subfigure}%
\begin{subfigure}{.5\textwidth}
  \hspace{-0.3cm}
  \includegraphics[width=10cm,height=7cm]{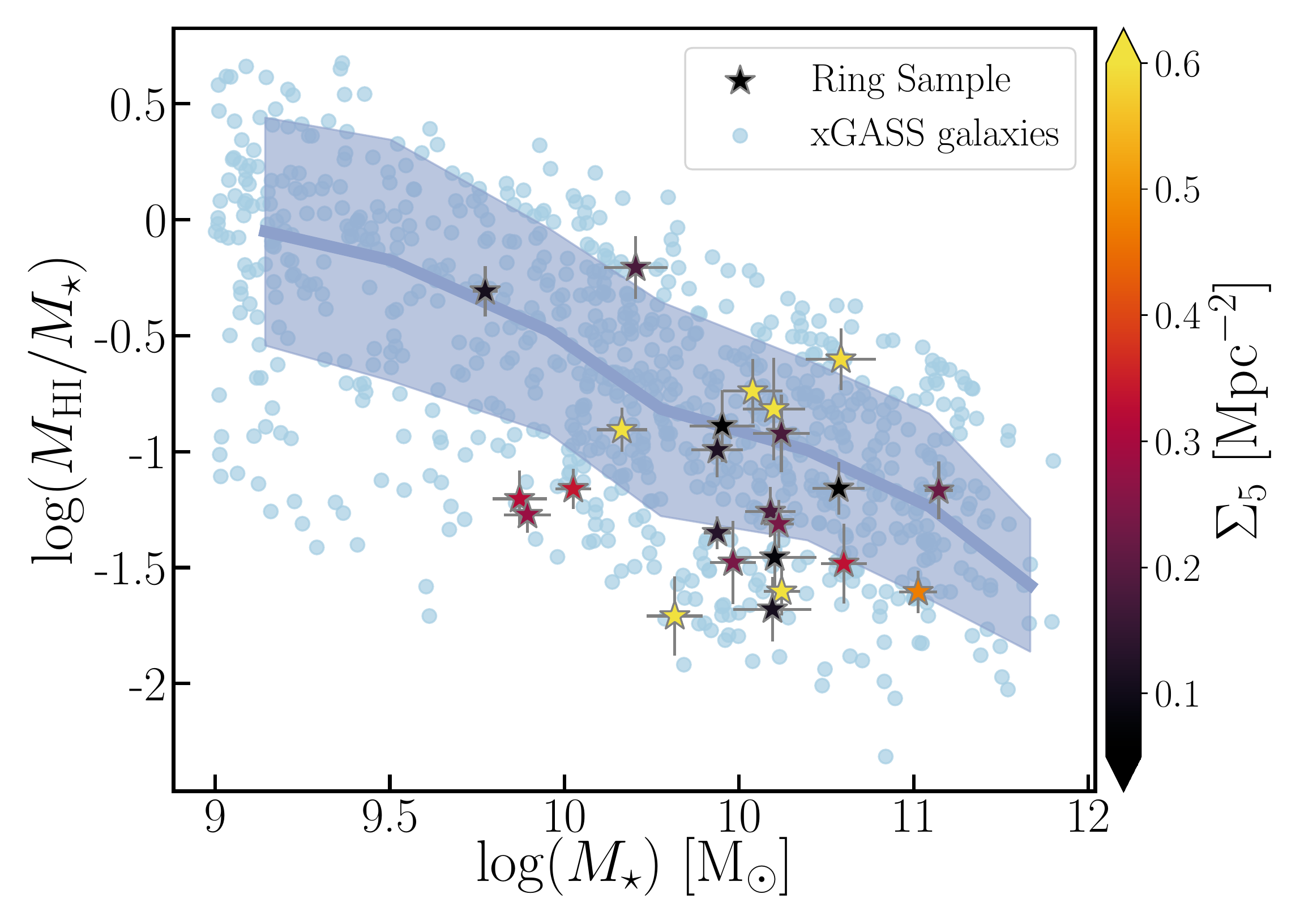}
  \caption{}
  \label{fig:gasfraction_Sig5}
\end{subfigure}
\caption{(a) The \h1 gas fraction vs the stellar mass of the ring galaxies compared with the xGASS galaxies (blue points). The green stars represent galaxies in our sample with a weak or no bar and orange stars represent galaxies with a strong bar (b) The M$_{\small \textrm{\h1}} -$ M$_{\star}$ relation for the rings sample, this time with the galaxies colour-coded in terms of their $\Sigma_{5}$ local environment density values. The dark purple line traces the rolling median of the xGASS galaxies with the shaded region representing the 1$\sigma$ scatter.}
\label{fig:gasfraction}
\end{figure*}

Fig~\ref{fig:gasfraction_a} shows the \h1 gas fraction plotted against the stellar mass for the ring galaxy sample ($f_{\textrm{\h1}} - M_{\star}$ relation). As mentioned, we segregate our ring galaxy sample into two -- those without bars and/or having a weak bar (green stars) and those observed to have a strong bar (orange stars). We compare our ring galaxy sample with the representative xGASS sample \citep{Catinella2018} (blue points) in order to infer how the \h1 gas fraction of the ring galaxies relates to their stellar mass. We observe that 11 ($\sim 46$\%) of our sample galaxies appear to be \h1-deficient compared to the xGASS sample (i.e., lying below the $1\sigma$ scatter about the running median of the xGASS sample). In addition, we find that 9 of the 11 ($\sim 82$\%) galaxies identified to be \h1-deficient are observed to have a strong bar. 

We quantify the degree of \h1-deficiency of the barred and unbarred galaxies in our sample using the xGASS galaxies as a representative sample. We compute the \h1 deficiency as follows \\

DEF$_{\textrm{\h1}} = \log(M_{\textrm{\h1,exp}}) - \log(M_{\textrm{\h1,obs}})$ \\

Where $\log(M_{\textrm{\h1,exp}})$ is the expected \h1 mass obtained from the rolling median of the $M_{\textrm{\h1}} - M_{\star}$ scaling relation for the xGASS sample, and $\log(M_{\textrm{\h1,obs}})$ is the observed \h1 mass calculated from our ATCA observations. We also measure the \h1 deficiency values for the sample by replacing the ATCA-derived \h1 mass with their HIPASS-derived \h1 mass. Higher the DEF$_{\textrm{\h1}}$ value, more \h1-deficient the galaxy and vice-a-versa. In Table~\ref{tab:HI-deficiency} we show the computed median \h1 deficiency values for both the unbarred and barred galaxies in our sample, and estimate the standard error on the computed median values using a bootstrap re-sampling method. We find that the barred galaxies are significantly more \h1-deficient compared to the unbarred galaxies in our sample. 

\begin{table}
\caption{The median \h1 deficiency values computed for the unbarred and barred galaxies in our sample, using both their ATCA- and HIPASS-derived \h1 masses. The standard error on the median value is quoted within the brackets.}
\label{tab:HI-deficiency}
\begin{tabular}{lcc}       
\hline \hline
 & DEF$_{\textrm{\h1}}$ (ATCA) & DEF$_{\textrm{\h1}}$ (HIPASS) \\
 &  &  \\
\hline         
Unbarred & -0.015 (0.131) & -0.087 (0.106) \\
\hline
Barred & 0.455 (0.154) & 0.298 (0.174) \\
\hline
\end{tabular}
\end{table}

\h1 deficiency in galaxies has been generally attributed to the effects of the environment. In dense environments, a number of processes tend to strip the \h1 gas from the discs of galaxies, thereby rendering them \h1-deficient. Such processes include, tidal interactions/stripping (\citealt{Byrd90};~\citealt{Rots90}), ram pressure stripping (\citealt{gunn72};~\citealt{Chung09}), harassment \citep{moore96} among others. To examine if environmental processes are severely affecting the \h1 gas properties of the observed \h1-deficient ring galaxies in our sample, we measure the local environment density for our sample galaxies. We make use of the 5th Nearest-Neighbour projected environment density measure, $\Sigma_{5}$ (Mpc$^{-2}$) ~\citep{baldry06}, to probe the local density of the environment the sample galaxies are residing in.

We follow the methods described in \cite{Murugeshan2020} to estimate the $\Sigma_{5}$ values for our sample. The values are estimated based on the 2MASS Redshift Survey (2MRS) catalogue \citep{huchra12}. With more than $\sim 43,000$ spectroscopic redshift measurements within the magnitude limits, $K_s \leq 11.75$ mag and | b | > 5\deg, the 2MRS catalogue is complete to 97.6\% and covers 91\% of the entire sky. The below steps are followed to measure the $\Sigma_{5}$ values: 

\begin{itemize}
    \item First, we make the 2MRS catalogue volume-limited. We make a velocity cut to the catalogue and include only galaxies with systemic velocities in the range $200 - 8000$ \kms. This is then followed by imposing an absolute $K_{s}$-band magnitude cut of $M_K < -23.45$, corresponding to the survey's limiting apparent magnitude of 11.75 mag at the highest velocity edge (8000 \kms), thus making the 2MRS sample volume-limited.
    \item The $\Sigma_5$ value is computed as $\Sigma_5 = 5/\pi D^2$, where $D$ is the projected distance in Mpc to the 5th nearest-neighbour within $\pm ~500$\kms of the target galaxy.
\end{itemize}

Fig.~\ref{fig:gasfraction_Sig5} shows the $f_{\textrm{\h1}} - M_{\star}$ relation for the ring sample, but now the galaxies are colour-coded based on their $\Sigma_5$ values. We find that overall, the majority of our sample galaxies are from lower-density environments such as the field and/or loose groups. However, few galaxies in our sample are from high-density environments such as dense groups and cluster.  Based on the $\Sigma_5$ values and the \h1 and optical morphology of the galaxies in our sample, we do not find any significant evidence that the \h1-deficiency is driven by environmental effects. This is suggestive that other internal processes may be driving their \h1-deficiency. As mentioned earlier, since $\sim 82$\% of the \h1 deficient galaxies in our sample possess a strong bar, it may be the case that the bar is playing an important role in driving the \h1 deficiency among the barred galaxies in our sample. 

The bar has been proposed to play a vital role in the evolution of galaxies, specifically in the context of angular momentum (AM) transport and redistribution of matter within the galactic disc (\citealt{Sorensen1976};~\citealt{Athanassoula1985};~\citealt{Knapen2005}). Previous simulation studies have shown that the gas in galactic discs experience torques due to the non-axisymmetric potential of the bar, leading to the dissipation of energy due to collisions in their orbit about the bar, consequently leading to loss in AM and funneling towards the centre of the galaxy (\citealt{Bournaud2002};~\citealt{Knapen2005b};~\citealt{Athanassoula2013}). In addition to driving gas to the centre within the co-rotation radius, the bar also effectively transfers AM to the outer parts of the disc, thereby increasing the stability of the \h1 gas in the outer disc, which remains unavailable for collapse to form stars \citep{Masters2012}. This duel effect i.e, funneling of the gas to the centre inducing star formation and exhaustion of the \h1 gas (\citealt{Lin2017}), and the transport of AM to the outer disc makes those galaxies having strong bars \h1 deficient and redder due to quenching (\citealt{Masters2012};~\citealt{Newnham2020};~\citealt{Fernandez2021}). Our sample galaxies are consistent with these observations, wherein we find that the majority of the strongly barred galaxies are more \h1-deficient compared to those galaxies with no bar or weakly barred. 

We note that there may be other potential processes driving the \h1-deficiency in galaxies other than the effects of the bar. This may include interaction driven tidal perturbations that may lead to starbursts (e.g.,~\citealt{Knapen2009};~\citealt{Bergvall2016}) and consequent depletion of the \h1 gas; effects of star formation-driven outflows and the Active Galactic Nuclei (AGN) in ionising and expelling some fraction of the accreted \h1 gas out of the disk and into the circumgalactic medium (e.g.,~\citealt{Debuhr2012};~\citealt{Harrison2014};~\citealt{Cheung2016}). All of these processes may be contributing to the observed \h1-deficiency in galaxies. However, as noted earlier in this section, a majority of the galaxies in our sample are from low to intermediate environments and so the effects of tidal interactions is likely to be minimal. In addition, only three galaxies in our sample (NGC 2217, NGC 1808 and NGC 6300) are identified to have any AGN activity. We do note that the current lack of nuclear activity in the majority of the galaxies in our sample does not necessarily mean the AGN was not turned on at an earlier time. This is because AGN activity in galaxies tend to turn on and off at shorter timescales $\sim 10^5$ yrs \citep{Schawinski2015}. Even with this caveat in mind, it is unlikely that AGN-driven winds have played a significant role in driving the \h1 deficiency among our sample of ring galaxies, mainly because the galaxies in our sample are typically high-mass i.e., $\log(M_{\star}/M_{\odot}) \geq 10$, while AGN-driven \h1 depletion is generally observed to impact low-mass galaxies i.e., $\log(M_{\star}/M_{\odot}) \leq 10$ (see for example \citealt{Bradford2018} and references therein). Thus, the majority of our sample is likely not severely affected by AGN-driven \h1 deficiency.  In the next section, we investigate the effects of the bar on the \h1 gas distribution among our sample galaxies, as well as how the bars may be pivotal in both funneling \h1 gas to the centres of galaxies as well as driving resonance rings in the inner and outer parts of the galactic disc. 

\begin{table*}
\caption{Ring galaxies derived sample properties. Col (1): Common name; Cols (2) - (3): Right Ascension and Declination in J2000; Col (4): Sytemic velocity; Col (5): Distance; Col (6): Integrated flux derived from SoFiA; Cols (7) - (8): \h1 and stellar mass; Col (9): 5th nearest-neighbour density estimate; Col (10): Whether the galaxy has a strong, weak bar or does not have a bar. We refer to The Carnegie-Irvine Galaxy Survey (CGS;~\citealt{Ho2011}) and the Third Reference Catalogue of Bright Galaxies \citep{devaucouleurs91} to confirm presence or absence of a bar.}
\label{tab:sample-properties}
\begin{tabular}{lccccccccc}       
\hline \hline
Name & RA & Dec & V$_{\textrm{sys}}$ & D & S$_{\textrm{int}}$ & $\log(M_{\textrm{\h1}})$ & $\log(M_{\star})$ & $\Sigma_5$ & Barred \\
 & [J2000] & [J2000] & [\kms] & [Mpc] & [Jy.\kms] & [M$_{\odot}$] & [M$_{\odot}$] & [Mpc$^{-2}$] & \\
(1)   & (2) & (3) & (4) & (5) & (6) & (7) & (8) & (9) & (10)\\
\hline         
ESO215-31	&	11h10m34.79s	&	-49d06m09.4s	&	2722	&	35.20	&	12.50	$\pm$	0.27	&	9.56	$\pm$	0.12	&	10.45	$\pm$	0.09	&	0.05 & Yes	\\
ESO269-57	&	13h10m04.43s	&	-46d26m14.4s	&	3106	&	43.24	&	38.5 $\pm$	0.81	&	10.19	$\pm$	0.10	&	10.79	$\pm$	0.10	&	0.59 & Yes	\\
NGC1326	&	03h23m56.40s	&	-36d27m52.8s	&	1360	&	14.95	&	7.67	$\pm$	0.26	&	8.61	$\pm$	0.11	&	10.32	$\pm$	0.08	&	1.59 & Yes	\\
NGC3358	&	10h43m33.02s	&	-36d24m38.5s	&	2988	&	38.30	&	6.00	$\pm$	0.33	&	9.32	$\pm$	0.15	&	10.80	$\pm$	0.07	&	0.33 & No	\\
IC5267	&	22h57m13.57s	&	-43d23m46.1s	&	1712	&	21.26	&	20.1	$\pm$	0.56	&	9.33	$\pm$	0.06	&	10.59	$\pm$	0.07	&	0.18 & No	\\
NGC1302	&	03h19m51.18s	&	-26d03m37.6s	&	1710	&	11.21	&	15.70	$\pm$	0.44	&	8.67	$\pm$	0.09	&	9.87	$\pm$	0.08	&	0.32 & Yes	\\
NGC1398	&	03h38m52.13s	&	-26d20m16.2s	&	1396	&	19.78	&	27.70	$\pm$	0.53	&	9.41	$\pm$	0.07	&	11.01	$\pm$	0.05	&	0.47 & Yes	\\
NGC2217	&	06h21m39.78s	&	-27d14m01.5s	&	1619	&	21.87	&	37.50	$\pm$	0.80	&	9.63	$\pm$	0.06	&	10.79	$\pm$	0.07	&	0.05 & Yes	\\
NGC7020	&	21h11m20.09s	&	-64d01m31.2s	&	3201	&	29.40	&	4.96	$\pm$	0.29	&	9.01	$\pm$	0.14	&	10.48	$\pm$	0.07	&	0.24 & No	\\
NGC1291	&	03h17m18.59s	&	-41d06m29.0s	&	839	&	4.40	&	91.60	$\pm$	1.09	&	8.62	$\pm$	0.03	&	9.89	$\pm$	0.07	&	0.28 & Yes	\\
NGC1371	&	03h35m01.34s	&	-24d55m59.6s	&	1463	&	23.79	&	45.50	$\pm$	1.01	&	9.78	$\pm$	0.18	&	10.60	$\pm$	0.09	&	0.7	 & Weak \\
ESO179-IG013	&	16h47m20.05s	&	-57d26m24.6s	&	842	&	10.90	&	104.00	$\pm$	1.44	&	9.46	$\pm$	0.09	&	9.77	$\pm$	0.04	&	0.11 & No \\
NGC1079	&	02h43m44.34s	&	-29d00m12.1s	&	1452	&	25.20	&	18.60	$\pm$	0.63	&	9.45	$\pm$	0.08	&	10.44	$\pm$	0.07	&	0.12 & Weak	\\
NGC5101	&	13h21m46.24s	&	-27d25m49.9s	&	1868	&	17.23	&	20.00	$\pm$	0.81	&	9.15	$\pm$	0.06	&	10.60	$\pm$	0.12	&	0.08 & Yes	\\
IC5240	&	22h41m52.38s	&	-44d46m01.8s	&	1765	&	25.37	&	13.20	$\pm$	0.33	&	9.30	$\pm$	0.08	&	10.61	$\pm$	0.04	&	0.24 & Yes	\\
NGC1350	&	03h31m08.12s	&	-33d37m43.1s	&	1905	&	18.83	&	12.50	$\pm$	0.31	&	9.02	$\pm$	0.08	&	10.62	$\pm$	0.05	&	1.07 & Yes	\\
NGC2369	&	07h16m37.73s	&	-62d20m37.4s	&	3240	&	35.30	&	17.00	$\pm$	0.56	&	9.70	$\pm$	0.13	&	10.62	$\pm$	0.08	&	0.18 & No	\\
NGC7098	&	21h44m16.12s	&	-75d06m40.8s	&	2381	&	29.10	&	40.20	$\pm$	1.15	&	9.90	$\pm$	0.10	&	11.07	$\pm$	0.04	&	0.22 & Weak	\\
NGC1543	&	04h12m43.25s	&	-57d44m16.7s	&	1176	&	17.19	&	11.80	$\pm$	0.38	&	8.92	$\pm$	0.07	&	10.60	$\pm$	0.11	&	0.11 & Yes	\\
NGC1533	&	04h09m51.84s	&	-56d07m06.4s	&	790	&	20.20	&	65.54	$\pm$	2.31	&	9.80	$\pm$	0.08	&	10.54	$\pm$	0.09	&	1.59 & Yes	\\
NGC1808	&	05h07m42.34s	&	-37d30m47.0s	&	995	&	9.29	&	89.05	$\pm$	0.89	&	9.26	$\pm$	0.05	&	10.16	$\pm$	0.07	&	5.64 & No	\\
NGC7531	&	23h14m48.50s	&	-43d35m59.8s	&	1596	&	22.22	&	85.44	$\pm$	1.63	&	10.00	$\pm$	0.10	&	10.20	$\pm$	0.09	&	0.18 & No	\\
NGC1433	&	03h42m01.55s	&	-47d13m19.5s	&	1076	&	9.47	&	34.60	$\pm$	0.61	&	8.86	$\pm$	0.07	&	10.03	$\pm$	0.05	&	0.34 & Yes	\\
NGC6300	&	17h16m59.47s	&	-62d49m14.0s	&	1109	&	12.26	&	34.50	$\pm$	0.52	&	9.09	$\pm$	0.06	&	10.44	$\pm$	0.04	&	0.13 & Yes	\\
\hline
\end{tabular}
\end{table*}

\section{Bar-driven resonance rings}
\label{sec:Bar_driven_resonance_rings}

In this section, we study the effects of bars in driving the ring structure among the secularly evolving ring galaxies in our sample. There is a lot of evidence from both observations and models that suggest that bars can lead to the formation of resonance rings (\citealt{Schwarz1981};~\citealt{Buta1986};~\citealt{Buta1991};~\citealt{Sellwood1993}). The non-axisymmetric gravitational potential of a bar can trigger density waves that leads to orbital resonances where gas and stars can accumulate to form ring-like structures. The location of these orbital resonances (also called Lindblad resonances) depend primarily on the bar patter speed (\omegabar), the angular velocity ($\Omega$) and the radial epicyclic frequency ($\kappa$) of the gas and stars orbiting the galaxy (\citealt{Lindblad1964};~\citealt{Buta99}). The resonances occur at the radius where the following condition is met: \\

\omegabar$ = \Omega \pm \kappa/m$ \\

where $m$ is an integer. The most commonly observed resonances in galaxies are the Outer Lindblad Resonance (OLR), with \omegabar$ = \Omega + \kappa/2$, the Inner Lindblad Resonance (ILR), where \omegabar = $ \Omega - \kappa/2$, the inner Ultra-harmonic Resonance (iUHR) \omegabar$ = \Omega - \kappa/4$ and the outer Ultra-harmonic Resonance (oUHR), where \omegabar$ = \Omega + \kappa/4$. 

We measure the location of the various resonance radii based on the semi-major axis radius ($a_{\textrm{bar}}$) and the angular pattern speed of the bar (\omegabar).  Various methods have been proposed in the literature to measure the length of bars in galaxies including visual methods based on where the isophotes diverge from the typical boxy shape of the bar (e.g.,~\citealt{Garcia-Barreto2022}); bar radius defined as the radius at which the intensity along the major axis of the bar drops to half the intensity of the peak \citep{Salo2015}; as well as based on fourier analysis (\citealt{Laurikainen2002};~\citealt{Buta2006}).

\begin{figure}
    \centering
    \includegraphics[width=1\columnwidth]{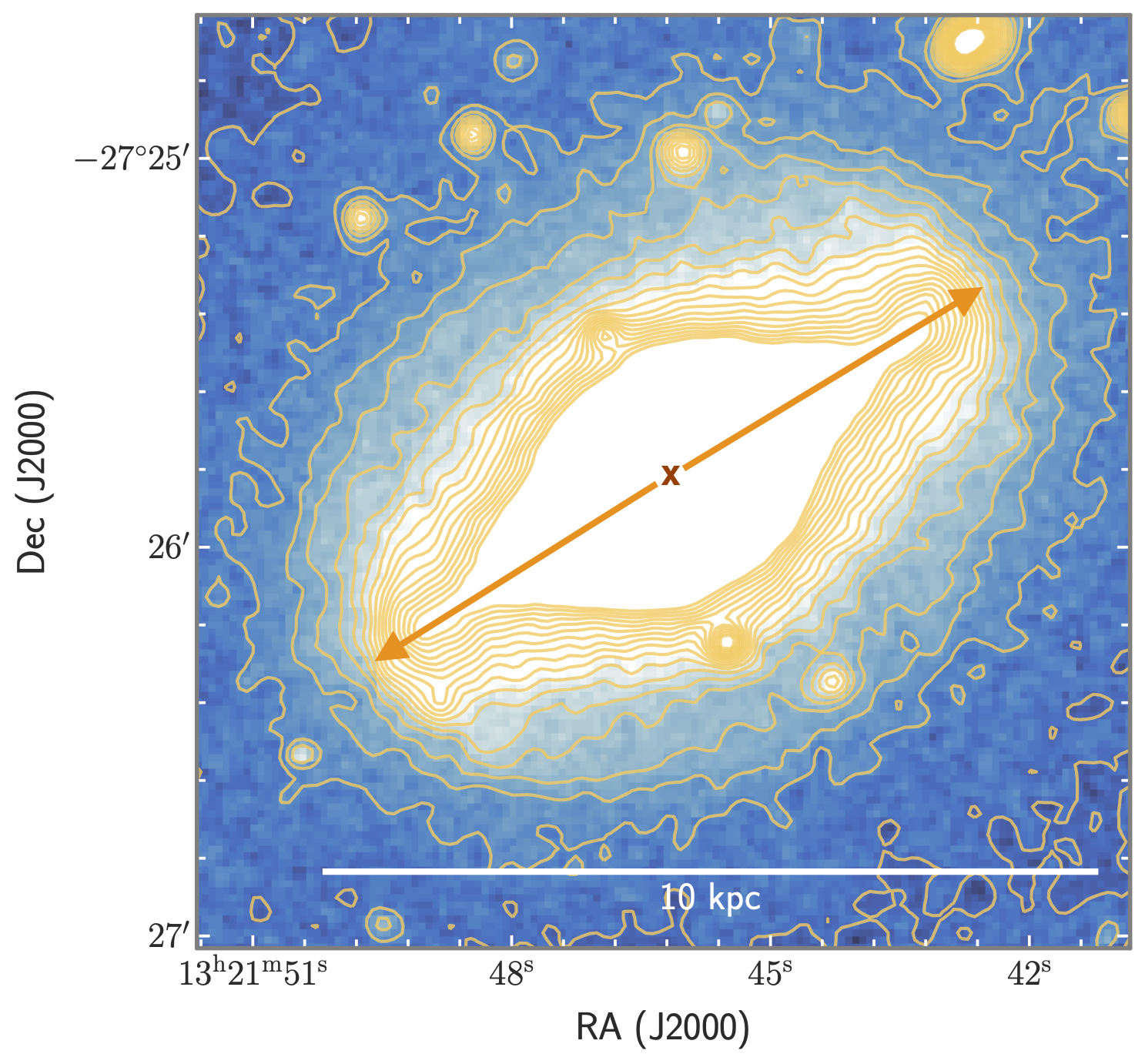}
    \caption{DSS IR-band image of the zoomed-in central region of the ring galaxy NGC 5101, showing the bar. The yellow curves are the IR-band isophotal contours overlaid on the image. The red cross denotes the centre of the galaxy and the orange arrows represent the length of bar semi-major axis radius towards each end of the bar along the major axis of the bar.}
    \label{fig:NGC5101_Bar_isophotes}
\end{figure}

We measure the bar size visually following the methods described in \cite{Garcia-Barreto2022}. As the stellar population of the bar is mostly dominated by older stars which are better represented in the IR (see for example~\citealt{Buta99};~\citealt{Eskridge2000};~\citealt{Herrera-Endoqui2015} and references therein), we make use of the DSS IR-band image to visually determine the radius of the bar ($a_{\textrm{bar}}$) for our sample of barred ring galaxies. This involves plotting isophotal contours on the DSS IR image and defining the bar radius as the distance from the centre of the galaxy to the last isophote that follows a boxy contour (see Fig.~\ref{fig:NGC5101_Bar_isophotes}). We measure the length of the semi-major axis radius of the bar from the centre to each ends of the isophote along the major axis of the bar and take the bar radius to be the average of the two, we term this value  $a_{\textrm{bar}}^{\textrm{uncorr}}$. We compute the standard error on the average value of $a_{\textrm{bar}}^{\textrm{uncorr}}$ and propagate this error when computing the de-projected bar radius $a_{\textrm{bar}}$.

The measured average uncorrected bar radius ($a_{\textrm{bar}}^{\textrm{uncorr}}$) is then de-projected to the kinematical major axis of the galaxy, to obtain the corrected bar radius ($a_{\textrm{bar}}$). Again, this is done following the method described in \citet{Garcia-Barreto2022} as: \\

$a_{\textrm{bar}} = a_{\textrm{bar}}^{\textrm{uncorr}} \sqrt{\cos ^2(\phi)+\left(\sin ^2(\phi) / \sin ^2(i)\right)}$. \\

Here $\phi$ is the absolute difference between the median kinematic position angle (P.A) of the disk of the galaxy and the position angle of the bar and $i$ is the inclination angle of the optical disc of the galaxy as described in Section~\ref{subsec:3D_modelling}. $a_{\textrm{bar}}$ is then converted to a physical size in units of kpc using the distance ($D$) to the source. We propagate the errors associated with $a_{\textrm{bar}}^{\textrm{uncorr}}$, $\phi$ and assume a nominal 10\% error on the distance measurements to compute the errors on $a_{\textrm{bar}}$. We list the $a_{\textrm{bar}}$ values in Table~\ref{tab:bar_analysis}.

Once we determine the corrected/de-projected bar radius, we assume the ratio of co-rotation radius of the bar and bar radius, $R \equiv R_{\textrm{CR}}/a_{\textrm{bar}} = 1$, so that $R_{\textrm{CR}} = a_{\textrm{bar}}$. A number of previous works have shown that the value of $R$ ranges from 1 to 1.6 (e.g.,~\citealt{Debattista2000};~\citealt{Guo2019};~\citealt{Schmidt2019}). While other works have shown that $\sim 90$\% of the barred galaxies have $1.0 \leq R \leq 1.4$ (~\citealt{Athanassoula1992};~\citealt{Debattista2000};~\citealt{Starkman2018}). Therefore, setting $R = 1$ is a reasonable approximation to make for our analysis. Indeed, we find that setting $R = 1.2$, which is the average of the range of reported values does not significantly affect our results.
    
Next, we derive the angular velocity profile (\omegaHI) from the \h1 gas for our sample galaxies using the \h1 rotation curve ($v_{\textrm{rot}}(r)$) extracted from 3DBarolo (see Section~\ref{subsec:3D_modelling}). The observed rotation curve is fit with a simple analytic function introduced in previous works (e.g.,~\citealt{Boissier2003};~\citealt{Obreschkow2016}) and of the form, \\

\begin{figure*}
\hspace{-1.9cm}
    \includegraphics[width=1.1\columnwidth]{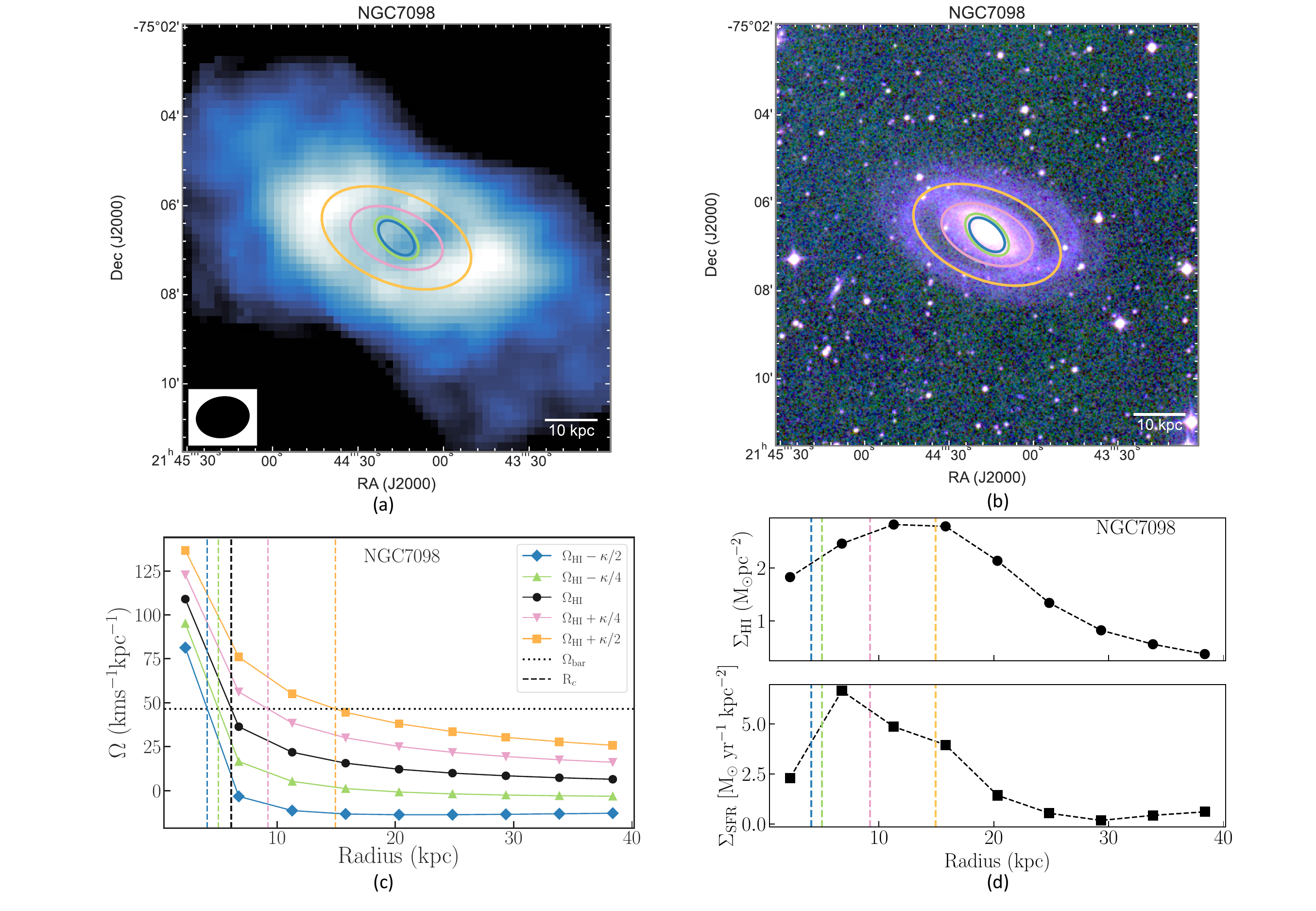}
    \caption{(a) \h1 moment 0 intensity map of NGC 7098. The synthesised beam is shown in the left bottom corner. (b) DSS composite optical image of NGC 7098. The blue, green, pink and orange ellipses represent the location of the ILR, iUHR, oUHR and OLR, respectively. (c) The plot shows the angular velocity (\omegaHI;black line) derived from the rotation velocity of the \h1 gas as a function of radius for the galaxy NGC 7098. The blue, green, pink and orange profiles represent the ILR, iUHR, oUHR and OLR Lindblad resonance curves respectively. The vertical black dashed line represents the measured bar co-rotation radius (R$_c$), the horizontal black dotted line represents the bar pattern speed (\omegabar). The blue, green, pink and orange vertical dashed lines denote the location of the ILR, iUHR, oUHR and OLR resonance radii respectively. (d) Top panel: Radial profile of the \h1 surface density of the galaxy NGC 7098. Bottom panel: Radial profile of the SFR surface density.}
    \label{fig:NGC7098_resonance_mosaic}
\end{figure*}

$v_{\textrm{rot}}(r) = v_{\max}\left(1-e^{-r / r_{\text{max}}}\right)$, \\

where $v_{\max}$ is the maximum rotation velocity corresponding to the flat part of the rotation curve and $r_{\text{max}}$ is the radius where the rotation curve transitions to the flat part. Both $v_{\max}$ and $r_{\text{max}}$ are left as free parameters during the fit. We use the best fitting function describing the rotation curve to then derive the angular velocity profile of the \h1 gas, \omegaHI $= v_{\textrm{rot}}(r)/r$, where $v_{\textrm{rot}}(r)$ is the circular rotation velocity in \kms and $r$ is the radius in kpc. We note that the choice of the analytic function used to fit the rotation curves was motivated on the basis of its simplicity as well as the fact that we were able to obtain reasonably good fits for our sample. Following this we measure the bar pattern speed, \omegabar. \omegabar is the angular velocity corresponding to the bar co-rotation radius $R_{\textrm{CR}}$ (see panel (c) in Fig.~\ref{fig:NGC7098_resonance_mosaic}). To determine the various Lindblad resonance radii, we first need to derive the epicyclic frequency ($\kappa$), which can be approximated as follows: \\

$\kappa(r)=\sqrt{4 \Omega_{\textrm{\h1}}^2+2 r \Omega_{\textrm{\h1}}\left(d \Omega_{\textrm{\h1}} / d r \right)}$ \\

Panel (c) of Fig.~\ref{fig:NGC7098_resonance_mosaic} shows a plot of the angular velocity profile \omegaHI as well as the Lindblad curves, \ILR, \iUHR, \oUHR and \OLR for the galaxy NGC 7098 (black, blue, green, pink and orange curves respectively). The location of the co-rotation radius ($R_{\textrm{CR}}$) of the bar is represented by the black dashed vertical line in panel (c) of Fig.~\ref{fig:NGC7098_resonance_mosaic}. The bar pattern speed (\omegabar) is represented by the horizontal black dotted line. The ILR, iUHR, oUHR and OLR radii are determined as the radius at which \omegabar intersects the \ILR, \iUHR, \oUHR and \OLR curves respectively. The position angle of the inner rings (ILR and iUHR) is set to be equal to that of the bar position angle, as many previous studies have shown that the inner rings are typically aligned parallel to the major axis of the bar (e.g.,~\citealt{Buta1996}). The position angle of the outer rings (oUHR and OLR) are set equal to the median kinematic P.A of the galaxy. 

Panels (a) and (b) in Fig.~\ref{fig:NGC7098_resonance_mosaic} show the \h1 moment 0 map and the optical DSS image respectively of NGC 7098 with the ILR, iUHR, oUHR and OLR ring locations indicated by the blue, green, pink and orange ellipses. NGC 7098 is observed to posses an inner ring as well as an outer ring (see DSS optical image in panel (b) of Fig.~\ref{fig:NGC7098_resonance_mosaic}) and we find that the predicted location of the oUHR (pink ellipse) coincides nicely with the observed inner ring and the predicted OLR (orange ellipse) traces the outer ring in very good agreement. Additionally, we also observe an \h1 ring located at the OLR radius. 

\begin{table*}
\caption{Properties of the bar and Lindblad resonances. Col (1): Common name; Col (2): Average length of the semi-major axis of the bar in kpc; Col (3): Bar patter speed in units of \kms kpc$^{-1}$; Cols (4) -- (7): measured radius of the inner Lindblad resonance (ILR), inner Ultra-harmonic resonance (iUHR), outer Ultra-harmonic resonance (oUHR) and outer Lindblad resonance (OLR) respectively in units of kpc}
\label{tab:bar_analysis}
\begin{tabular}{lcccccc}       
\hline \hline
Name & $a_{\textrm{bar}}$ & \omegabar & R$_{\textrm{ILR}}$ & R$_{\textrm{iUHR}}$ & R$_{\textrm{oUHR}}$ & R$_{\textrm{OLR}}$ \\
 & [kpc] & [\kms kpc$^{-1}$] & [kpc] & [kpc] & [kpc] & [kpc]  \\
(1)   & (2) & (3) & (4) & (5) & (6) & (7) \\
\hline         
ESO215-31	&	10.6 $\pm$ 1.1	&	10 $\pm$ 1	&	NA	&	6.0	&	22.5	&	NA	\\
NGC1326	&	3.2 $\pm$ 0.3	&	52 $\pm$ 5	&	1.91	&	2.5	&	4.6	&	6.8	\\
NGC1398	&	5.2 $\pm$ 0.5	&	47 $\pm$ 3	&	NA	&	NA	&	11.4	&	21.5	\\
NGC2217	&	7.5 $\pm$ 0.8	&	27 $\pm$ 2	&	3.4	&	4.6	&	13.8	&	25.1	\\
NGC1371	&	3.9 $\pm$ 0.4	&	57$\pm$ 3	&	NA	&	NA	&	6.9	&	11.0	\\
NGC1079	&	4.3 $\pm$ 0.4	&	26 $\pm$ 1	&	NA	&	NA	&	13.0	&	NA	\\
NGC5101	&	6.7 $\pm$ 0.7	&	31 $\pm$ 3	&	NA	&	NA	&	12.0	&	NA	\\
IC5240	&	3.9 $\pm$ 0.4	&	44 $\pm$ 4	&	2.2	&	3.0	&	6.0	&	9.6	\\
NGC1350	&	5.3 $\pm$ 0.6	&	39 $\pm$ 4	&	NA	&	3.2	&	9.2	&	15.7	\\
NGC7098	&	6.1 $\pm$ 0.6	&	47 $\pm$ 10	&	4.1	&	5.1	&	9.2	&	14.9	\\
NGC1533	&	2.7 $\pm$ 0.3	&	32 $\pm$ 1	&	NA	&	NA	&	5.8	&	10.2	\\
NGC6300	&	2.2 $\pm$ 0.3	&	90 $\pm$ 13	&	NA	&	NA	&	2.6	&	3.6	\\
\hline
\end{tabular}
\end{table*}

In addition, we derive the star formation rate (SFR) surface density profiles for our sample to examine if the SFR is enhanced at the locations of the resonance rings. A number of previous works have highlighted that star formation is triggered at the resonance locations in the disc where the \h1 gas accumulates. We derive the SFR surface density profiles for the galaxies in our sample using their WISE $W_3$ and $W_1$ images procured from the NASA/IPAC IRSA\footnote{\url{https://irsa.ipac.caltech.edu/applications/wise/}} image cutout service. The $W_3$ flux is shown to be a good tracer of the SFR in galaxies, however, this flux is also contaminated by the contribution from older stellar population (~\citealt{Jarrett2011};~\citealt{Cluver17}). We correct for this following the methods described in \cite{Cluver17} and subtract $\sim 15.8\%$ of the $W_1$-band flux from the $W_3$ flux.

We first process both the $W_3$ and $W_1$ images to remove foreground stars. We run the source finding code SExtractor~\citep{SExtractor} to identify point-like sources in the images, following this we use the IRAF \citep{Tody1986} task \texttt{imedit} to replace the bad pixels of the star with random pixels surrounding the star. In the next step we place an annulus around each galaxy and measure the median local sky background values from both the $W_3$ and $W_1$ images, before subtracting this from the respective images. Next, we measure both the $W_3$ and $W_1$ flux within concentric ellipses, with position angle set to the median kinematic position angle of the galaxy, and convert the corrected $W_3$ flux to SFR following the methods described in \cite{Cluver17}.

In Panel (d) of Fig.~\ref{fig:NGC7098_resonance_mosaic} we plot both the azimuthally averaged \h1 surface density ($\Sigma_{\textrm{\h1}}$) and the SFR surface density profile ($\Sigma_{\textrm{SFR}}$) for NGC 7098. The inner region (< 10 kpc) of this galaxy is observed to have a dip in \h1 surface density ($\Sigma_{\textrm{\h1}}$), while showing a peak in SFR surface density. At the location of the resonance radii, there is an enhancement in $\Sigma_{\textrm{\h1}}$ and $\Sigma_{\textrm{SFR}}$. This is also observed in most of the barred galaxies in our sample (see \ref{appendix:notes_on_galaxies} for more details) and is consistent with the theory that bars drive resonances, where the disk stars and \h1 gas accumulate, and the gas eventually collapses to form new stars, thereby resulting in increased SFR at the resonance ring locations (\citealt{Buta1996}). Such ring-like structures are likely to remain in equilibrium for a few Gyrs as the resonant locations are regions in the galactic disc where the net torque is zero (e.g.,~\citealt{Buta99};~\citealt{Bournaud2005}).

We perform the above resonance model analysis for the rest of the barred galaxies in our sample to see if we are able to predict the location of the resonance rings consistently. We report that there is good agreement between the predicted locations of the resonances and the presence of actual ring-like structures (in \h1 and/or in the optical) for the majority of the galaxies. We note that depending on the angular velocity profiles, bar radius and pattern speed of the individual galaxies, we are unable to always measure all four resonances. We list the measured bar radius ($a_{\textrm{bar}}$), pattern speed (\omegabar) and various resonance radii for the sample in Table~\ref{tab:bar_analysis}. Figs.~\ref{fig:resonance_rings_appendix1} -- \ref{fig:resonance_rings_appendix5} in \ref{appendix:notes_on_galaxies} show the outcomes of the resonance model analysis for the other barred galaxies in our sample.

The measured values of the bar pattern speed (\omegabar) for our sample range from $\sim$ 10 -- 90 \kms kpc$^{-1}$, with a median value of $\sim 44$ \kms kpc$^{-1}$. This range is consistent with previous works that have measured \omegabar values in galaxies (e.g.,~\citealt{Corsini2011};~\citealt{Williams2021}). In addition, we found a few galaxies in our sample for which the \omegabar measurements were available in the literature and find that overall the values are in good agreement. For example, \cite{Moore1995} use VLA \h1 observations to determine \omegabar of NGC 1398 and estimate a value $\sim 46$ \kms kpc$^{-1}$, which is in very good agreement with our measured value of $47\pm3$ \kms kpc$^{-1}$. \cite{Garcia-Barreto1991} measure a pattern speed value of $\sim 60$ \kms kpc$^{-1}$ for NGC 1326, while we measure a pattern speed of $52\pm5$ \kms kpc$^{-1}$. We measure \omegabar$90\pm13$ \kms kpc$^{-1}$ for NGC 6300, which is in disagreement with the value measured by \cite{Ryder1996}, who estimate a bar pattern speed of $27 \pm 8$ \kms kpc$^{-1}$. However, we emphasise that the higher \omegabar value we measure predicts the existence of an OLR at a radius of $\sim 3.57$ kpc, which coincides with the location of both an \h1 and stellar ring in NGC 6300 (see Fig.~\ref{fig:resonance_rings_appendix3} in \ref{appendix:notes_on_galaxies}). While \cite{Ryder1996} report that assuming the lower bar pattern speed of $\sim 27$ \kms kpc$^{-1}$ would result in the location of an OLR at radius between 8 -- 15 kpc, where they find no \h1 or stellar ring. The fact that our \omegabar measurement for NGC 6300 better matches the resonance model instills confidence in our analysis.

Overall, from our analysis, we find that the bar driven resonance theory explains the location of the inner and outer rings that we observe among galaxies in our sample. It is indeed promising to note that we are able to reliably measure the location of the various Lindblad resonances using just the \h1 rotation curve and a visual measurement of the size of the bar. These results come in support of the important role played by the bar in not only funneling gas to the centers of galaxies but also in driving resonances and redistributing the angular momentum in galactic discs. As noted in Section~\ref{subsec:Sample_overview}, where we observed strongly barred galaxies in our sample to be in general more \h1-deficient compared to the weak and non-barred galaxies, it seems evident that bars accelerate the evolution of galaxies via their ability to both use-up large reserves of the \h1 gas as well as making gas in the outskirts unavailable for star formation due to a kick in angular momentum, thus driving quenching in galaxies.

\section{Summary and Future}
\label{sec:Summary_Future}

We have introduced the new \h1 in Ring Galaxies Survey (\h1-RINGS), comprising of 24 ring galaxies for which we have obtained high-resolution \h1 observations using the Australia Telescope Compact Array. The ring galaxies were selected on the basis of their large angular size, their declination range and a detection in the HIPASS catalogue. Our sample consists of a variety of ring galaxies including what have been regarded in the literature as resonance, collisional and interaction rings. 
We presented an overview of the sample, including their \h1, stellar and star formation properties. In addition, we also highlighted their \h1 morphology and kinematics. 

We segregated our sample of ring galaxies into two groups -- those with a strong bar and those with a weak or no bar. We find that in general the strongly barred galaxies are more \h1-deficient compared to the weakly barred or unbarred galaxies in our sample. We also probed the local environment densities of our sample galaxies in addition to looking at their optical and \h1 maps to examine if the environment is playing a prominent role in driving the \h1 deficiency in the galaxies, but find no strong evidence for the same. We therefore hypothesize as suggested by previous works (e.g.,~\citealt{Masters2012}) that the bars are potentially driving the observed \h1 deficiency, via their ability to funnel copious amounts of \h1 gas within the co-rotation radius of the bar to the centers of galaxies. In addition, the bars also transport angular momentum to the outer parts of the disc making the \h1 gas in the outskirts of galaxies gain a kick in angular momentum and remain stable against collapsing to form stars. This combined effect potentially leads to bar quenching in galaxies as noted in other works. 

We also employed Lindblad's resonance theory to test if we are able to predict the radius of the major resonances for a sub-sample of barred galaxies. To determine the resonance radii, we measured the bar radius based on a visual method and combined with the derived rotation curves of the galaxies, we measure the bar pattern speed (\omegabar) before computing the location of the ILR, iUHR, oUHR and OLR radius. The measured \omegabar values range from 10 -- 90 \kms kpc$^{-1}$ in good agreement with previous works. We find that the predicted location of the resonance radii are in good agreement with the observed physical locations of the \h1/optical inner and/or outer rings in our galaxies. Furthermore, we derived the azimuthally averaged \h1 ($\Sigma_{\textrm{\h1}}$) and SFR surface ($\Sigma_{\textrm{SFR}}$) density profiles for these galaxies and found that there is an enhancement in either $\Sigma_{\textrm{\h1}}$, $\Sigma_{\textrm{SFR}}$ or both corresponding to one or more of the resonances.

In forthcoming papers, we aim to study the global and resolved scaling relations for our sample of ring galaxies to see how the different types of ring galaxies behave in the various scaling relations. In addition, we also aim to examine the properties of the collisional and interaction galaxies in our sample. These are interesting systems as the ring structure is believed to be driven by external perturbations. Specifically, as part of the \h1-RINGS survey, we have high-resolution \h1 observations for a recently discovered collisional ring galaxy ESO179-IG013, also called ``Kathryn's Wheel'' \citep{Parker2015}, which is touted to be the nearest collisional ring galaxy and as such offers a great opportunity to probe its \h1 and star formation properties. In the near future we aim to address most of the questions posed in the introduction pertaining to the evolution of ring galaxies.

\begin{acknowledgement}
We would like to thank the anonymous referee for their feedback which significantly improved the quality of the paper. We would also like to extend our thanks to Genevieve Batten, Vaishali Parkash and George Heald for their useful comments and contributions. 

M.E.C. is a recipient of an Australian Research Council Future Fellowship (project No. FT170100273) funded by the Australian Government.

The Australia Telescope Compact Array is part of the Australia Telescope National Facility (https://ror.org/05qajvd42) which is funded by the Australian Government for operation as a National Facility managed by CSIRO. We acknowledge the Gomeroi people as the traditional owners of the Observatory site. This paper includes archived data obtained through the Australia Telescope Online Archive (http://atoa.atnf.csiro.au). This research has made use of the NASA/IPAC Ex- tragalactic Database (NED), which is operated by the Jet Propulsion Laboratory, California Institute of Technology, under contract with the National Aeronautics and Space Administration. This research has made use of the VizieR catalogue access tool, CDS, Strasbourg, France. The original description of the VizieR service was published in A\&AS 143, 23.This publication makes use of data products from the Two Micron All Sky Survey, which is a joint project of the University of Massachusetts and the Infrared Processing and Analysis Center/California Institute of Technology, funded by the National Aeronautics and Space Administration and the National Science Foundation. This publication makes use of data products from the Wide-field Infrared Survey Explorer, which is a joint project of the University of California, Los Angeles, and the Jet Propulsion Laboratory/California Institute of Technology, funded by the National Aeronautics and Space Administration. Part of this work was performed on the OzSTAR national facility at Swinburne University of Technology. OzSTAR is funded by Swinburne University of Technology and the National Collaborative Research Infrastructure Strategy (NCRIS). This work was written on the collaborative Overleaf platform \url{https://www.overleaf.com}. This research has made use of \texttt{python} \url{https://www.python.org} and python packages: \texttt{astropy} \citep{Astropy2013, Astropy2018}, \texttt{matplotlib} \url{http://matplotlib.org/} \citep{Hunter2007}, \texttt{APLpy} \url{https://aplpy.github.io/}, \texttt{pandas} \citep{Pandas}, \texttt{Jupyter notebook} \url{https://github.com/jupyter/notebook}, \texttt{NumPy} \url{http://www.numpy.org/} \citep{VanDerWalt2011}, \texttt{SciPy} \url{https://www.scipy.org/} \citep{Jones2001} and \texttt{CMasher} \url{https://github.com/1313e/CMasher} \citep{CMasher}. This research used Streamlit \url{https://streamlit.io/}.
\end{acknowledgement}

\bibliography{references}

\appendix

\section{Interactive web application to visualise the HI-RINGS sample}
\label{appendix:streamlit_app}

\begin{figure*}
    \centering
    \includegraphics[width=\columnwidth]{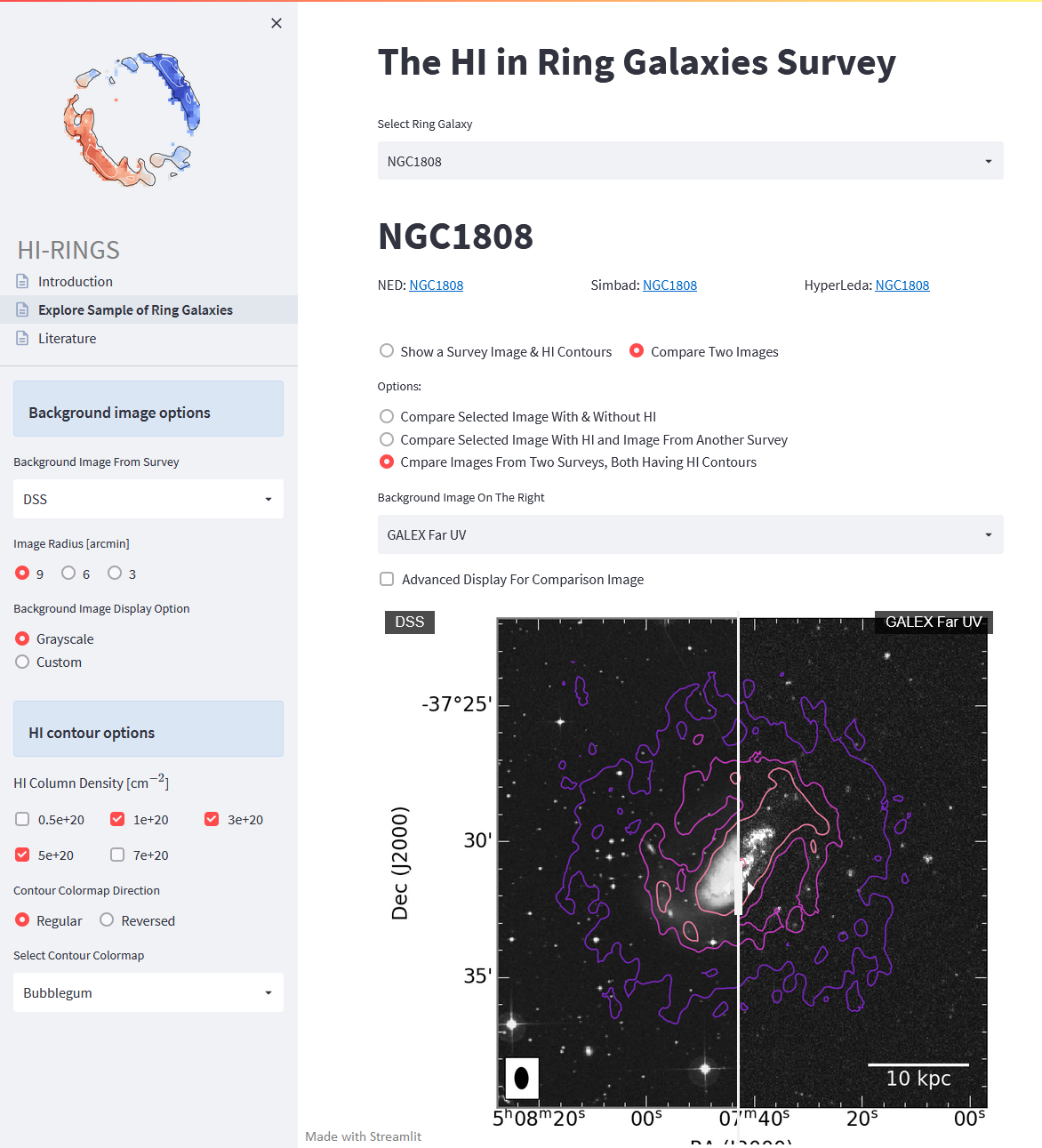}
    \caption{A screenshot showing the user interface of the \h1-RINGS Streamlit web application (\url{https://hi-rings.streamlitapp.com/}) that allows users to visualise our \h1-RINGS galaxy sample in a multi-wavelength approach. The main available options are: to select a ring galaxy, to show a survey image and \h1 contours and to compare two images. The user can select a background image from the listed surveys and adjust its display e.g., image radius, color - grayscale or custom colourmap and image stretch. The user can also adjust the \h1 contour display options, choosing which \h1 density contours they want to be displayed and which colourmap to be used for the displayed contours. These options are available for both a single image display and for a comparison of two images from two different surveys.}
    \label{fig:App}
\end{figure*}

Using Streamlit, an open-source Python library for turning data scripts into shareable web applications (\url{https://docs.streamlit.io/}), we have built a web application for interactive visualization of our \h1-RINGS survey:  \url{https://hi-rings.streamlitapp.com/}. In Fig.~\ref{fig:App} we show a screenshot of the web interface of the application. The application allows the user to interactively explore our sample of ring galaxies using \h1 intensity data in combination with images from the following surveys: DSS, DSS2 (Blue, Red and Infra-red) WISE (3.4$\mu$m, 4.6$\mu$m, 12$\mu$m and 24$\mu$m), 2MASS (J, H, K) and GALEX (Far-UV and Near-UV), obtained using the python library \texttt{astroquery} \citep{Astroquery2019}. There are two options for data display within the web application: 
\begin{itemize}
    \item \textbf{Display a survey image and \h1 contours}: this option is displayed upon opening the application. The user will see (by default) the \h1 contour at the $1\times 10^{20}$ cm$^{-2}$ column density level overlaid on top of the galaxy's DSS image, with an image radius of 9 arcmin. The user can then choose another galaxy, and/or adjust the display properties. 
    
    \item \textbf{Compare two images:} this option allows the user to visually compare two personally adjusted images of the selected ring galaxy by dragging the image slider from the centre of the images to the left and right. Image comparison is possible with and without overlaid \h1 contours, as well as comparison with different survey images in the background. Fig.~\ref{fig:App} shows these options (with \h1 contours), where the left side shows the DSS image while the right side displays the GALEX Far-UV image of the galaxy NGC 1808. 
\end{itemize}

In addition to selecting data display, it is possible to make further adjustments: image radius, display color options -- gray-scale or custom colour map and image stretch. Next, the user can adjust the \h1 contour display options, choosing which density contours they would like to display and which colour map scheme to be used for the contours. We believe such an interactive tool will be both informative as well as provide important visual context into the distribution of the \h1 and stellar components of the different types of ring galaxies in our sample. By allowing users to personally adjust displays we are extending data visualisation accessibility. Furthermore, a code of this web application is publicly available at: \url{https://github.com/ringgalaxies/HIRingGalaxies}. 
\textit{Note: As with any other application, it is highly likely that this application is going to be upgraded with time, and the user interface may differ to the current one shown in Fig.~\ref{fig:App}.}

\section{Notes on individual galaxies, 3DBarolo fit parameters and HI spectra}
\label{appendix:notes_on_galaxies}

In this section we specify notes on individual galaxies in our sample. As mentioned, the objective of the \h1-RINGS survey is to get a complete view of the properties of ring galaxies. Our sample include what appear to be resonance rings as well as collisional rings, ring galaxies with and without a bar. In addition, as mentioned earlier we also comment on galaxies that posses a bar and those that do not and probe the various mechanisms that may be driving their ring-like structure. We note that all morphological classes of the sample galaxies are based on the Carnegie atlas of galaxies \citep{Sandage1994}. We also discuss the location of the various Lindblad resonances for the barred galaxies in our sample using the resonance method that was described in Section~\ref{sec:Bar_driven_resonance_rings}. Additionally, we also note the location of the \h1 ring, optical ring, spiral arms and show the \h1 and SFR surface density profiles. 

Fig.~\ref{fig:parameter_plots1} and Fig.~\ref{fig:parameter_plots2} display the parameters obtained from the 3D fitting using 3DBarolo as described in Section~\ref{subsec:3D_modelling}. Each panel shows the rotation curve, the radial variation of the inclination ($i$) and position angles (PA) for the individual galaxies in our sample. In addition, Fig.~\ref{fig:spectra_1} and Fig.~\ref{fig:spectra_2} show the extracted ATCA and HIPASS spectra for our sample of ring galaxies. 

\begin{table*}
\caption{Ring galaxies in our sample categorised into those that are likely to be secularly evolving, those that are identified as collisional/interacting and also segregated on the basis of the presence or absence of bars. Individual notes on interesting systems is also made}
\label{Tab:Rings_classification}
\begin{tabular}{  l  p{8cm}  p{8cm} }
\toprule
    &  \textbf{Secular Rings}  & \textbf{Collisional/Interaction Rings} \\\midrule
\textbf{Barred}  

& Galaxies \textbf{ESO 215-31}, \textbf{NGC 1302}, \textbf{NGC 1398}, \textbf{ESO 269-57}, \textbf{NGC 1291}, \textbf{NGC 5101}, \textbf{NGC1543}, \textbf{NGC 1350}, \textbf{IC 5240}, \textbf{NGC 1433}  and \textbf{NGC 6300} are observed to show very similar optical morphologies. We find no strong evidence of interactions or external perturbations of the \h1 gas for these galaxies and therefore consider these galaxies to be following (semi-)secular evolution. All these galaxies are observed to posses a prominent bar, inner and outer (pseudo-)rings. We note that the \h1 disc of NGC 6300 is much more extended than its stellar disc unlike the other galaxies in this category, where the \h1 gas is observed to trace the outer ring and is not much more extended than the ring structure.

\vspace{0.3cm} For a subset of the barred galaxies for which we were able to derive accurate rotation curves, bar lengths and bar pattern speeds, we also test the resonance theory to predict the location of the various Lindblad resonances. Figs.~\ref{fig:resonance_rings_appendix1} -- \ref{fig:resonance_rings_appendix5} show all the relevant plots and images. It is observed that for all galaxies there is a good agreement between the predicted location of the resonances and the presence of ring-like structures either in the optical and/or the \h1. In addition, we also show that for most galaxies there is an enhancement in the azimuthally averaged SFR ($\Sigma_{\textrm{SFR}}$) and/or \h1 surface density ($\Sigma_{\textrm{\h1}}$) profiles corresponding to the radius of the resonances. 

& Galaxies \textbf{NGC 1326} and \textbf{NGC 2217} are observed to have \h1 tail-like extensions (towards the north for NGC 1326 and towards the north-east and south-west for NGC 2217, respectively, indicating some external perturbation and/or interaction). To what degree the external perturbation has affected the ring formation is uncertain. 

\vspace{0.3cm} \textbf{NGC 1533} -- while the optical morphology of this galaxy is observed to be that of an early-type, the \h1 gas distribution on the other hand is observed to be much more extended and is seen to be linked to two nearby smaller companion galaxies (IC 2038 and IC 2039) to the northwest. ~\cite{Ryan-Weber2004} note that the amount of extended \h1 gas in the tail leading to IC 2038/IC 2039 from NCG 1533 is significantly more than what is expected from the size and morphology of IC 2038/IC 2039. Thus, they conclude that the copious amounts of \h1 gas may have been stripped from a low surface brightness galaxy which is not visible in the optical images. Either way, it is certain that NGC 1533 has recently undergone an interaction and accredited the \h1 from a low surface brightness galaxy. The \h1 gas is observed to be distributed in the form a ring, with no optical counterpart.

\\\hline

\textbf{Unbarred}  
& Galaxies \textbf{NGC 7020}, \textbf{NGC 2369}, \textbf{NGC 7531} and \textbf{IC 5267} are observed to not possess bars. However, from their optical images we find that many of them possess an inner ring and all of them are observed to possess an outer (pseudo-)ring. 

\vspace{0.3cm} Galaxies \textbf{NGC 1079}, \textbf{NGC 1371} and \textbf{NGC 7098} are observed to have weak bars and outer rings. We note that the \h1 gas distribution in NGC 1371 is much more extended compared to its stellar disc.

\vspace{0.3cm} A Bright linear inclined disc structure is observed in \textbf{NGC 1808}, but not resembling a bar. The \h1 gas is seen to trace the narrow linear bright central stellar distribution as well as the outer pseudo-ring. 

\vspace{0.3cm} None of these galaxies show any significant signs of environmental effects as evidenced in their \h1 kinematic maps. Therefore, we attribute them to be secularly evolving. The origin of their rings therefore is not fully clear. 

& \textbf{NGC 3358} – This galaxy has a bright central nucleus and two tightly
wound spiral arms extend to form a ring-like outer structure. Corresponding to the outer optical ring is the \h1 ring of gas. The velocity map of this galaxy shows no obvious deviation from ordered rotational
motion. We note that this galaxy is part of an interacting system of at least seven galaxies part of the NGC 3347 group \citep{Makarov2011}. This galaxy does not appear to possess a bar, therefore the origin of the ring in this galaxy may be attributed to external tidal perturbations. 

\vspace{0.3cm} \textbf{ESO 179-IG013} -- This is an interesting candidate as it appears to have undergone an interaction based on both the optical and the \h1 maps (See upper right panels in Fig.~\ref{fig:mosaic2}) and is described as a collisional ring galaxy by \cite{Parker2015}, who term it ``Kathryn's Wheel". In terms of the galaxy morphology ESO 179-IG013 is classified as SB(s)m. The multiwavelength observations of ESO 179-IG013 by \cite{Parker2015} find a ring of star formation based on their H~$\alpha$ and optical observations (see Fig. 3 in their paper). 
While \cite{devaucouleurs91} classified this galaxy to posses a bar, \cite{Parker2015} argue that this elongated central structure may not actually be a bar but a residual late-type disc. In addition, \cite{Parker2015} identify the system to be the nearest collisional ring to the Milky Way with a systemic velocity of $\sim 840$~\kms. They also identify the smaller `bullet' galaxy that dropped through ESO 179-IG013, and find its star formation to be enhanced as well. In terms of the \h1 gas properties, this galaxy is considered the 60th brightest in the sample of 1000 brightest HIPASS galaxies \citep{Koribalski2004}, with an integrated flux of $\sim 104$Jy~\kms, corresponding to an estimated \h1 mass of $\log(\textrm{\h1}/M_{\odot}) \sim 9.46$ assuming a distance of 10.9 Mpc. A high-resolution \h1 follow-up of this galaxy has been lacking until now. As part of the \h1-RINGS survey, we present for the first time high-resolution \h1 follow-up observations for ESO 179-IG013. The \h1 velocity map of the system shows clear signs of perturbation with \h1 tails extending towards the south-east and north-west ends of the galaxy \\

\bottomrule
\end{tabular}
\end{table*}

\begin{landscape}
    \begin{figure}
    \centering
    \includegraphics[width=24cm,height=14cm]{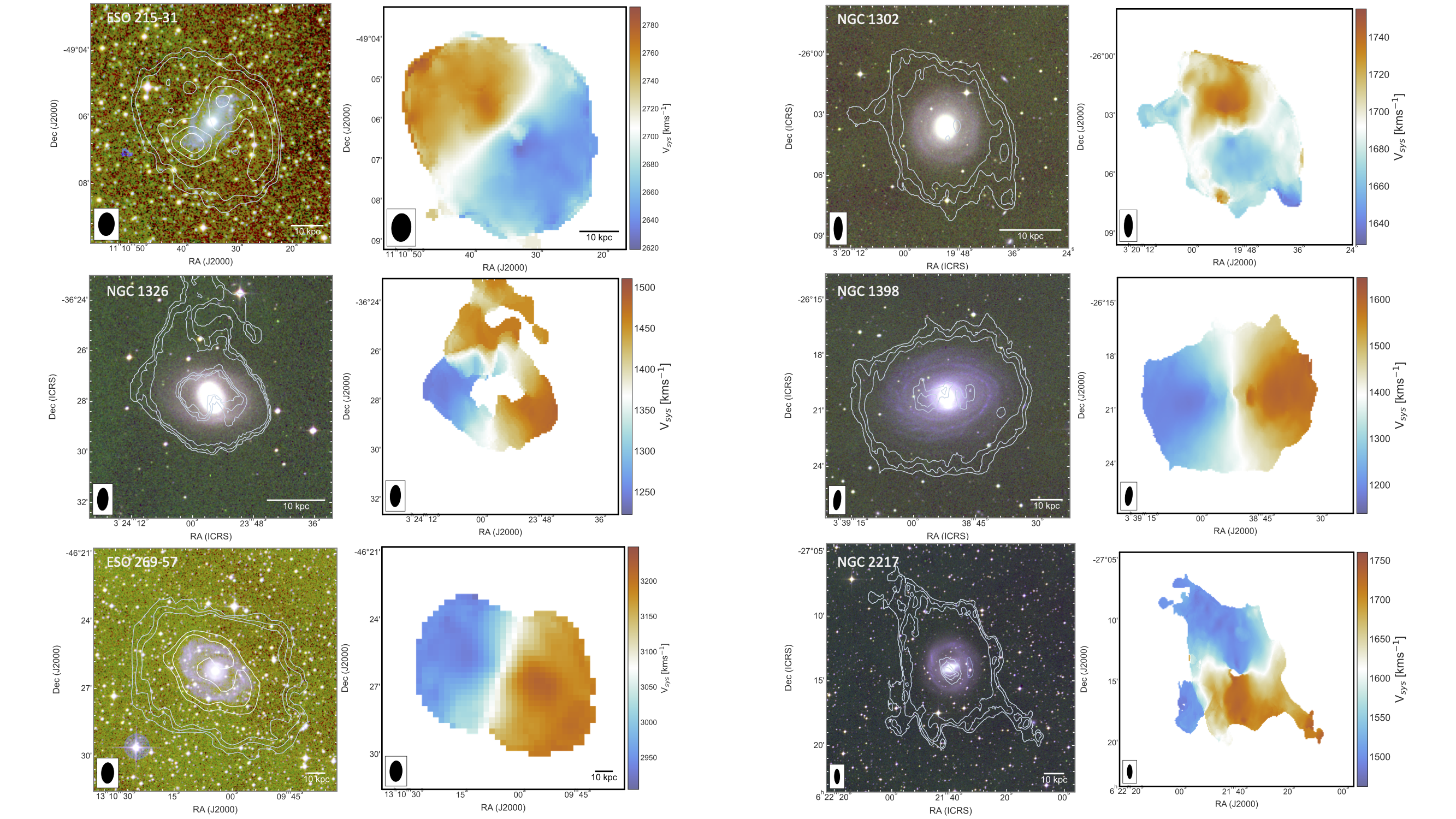}
    \caption{The panels to the left show the \h1 contours overlaid on top of the composite DSS images of the ring galaxies. The contour levels are at the N$_{\textrm{\h1}} \sim$ 0.1, 0.5, 1, 5 and 7$\times 10^{20}$ cm$^{-2}$. The panels to the right show the moment 1 (velocity) maps of the \h1 gas in the rings galaxies. The colour-bars show the velocity range for each individual galaxy.}
    \label{fig:mosaic1}
    \end{figure}
\end{landscape}

\begin{landscape}
    \begin{figure}
    \centering
    \includegraphics[width=24cm,height=14cm]{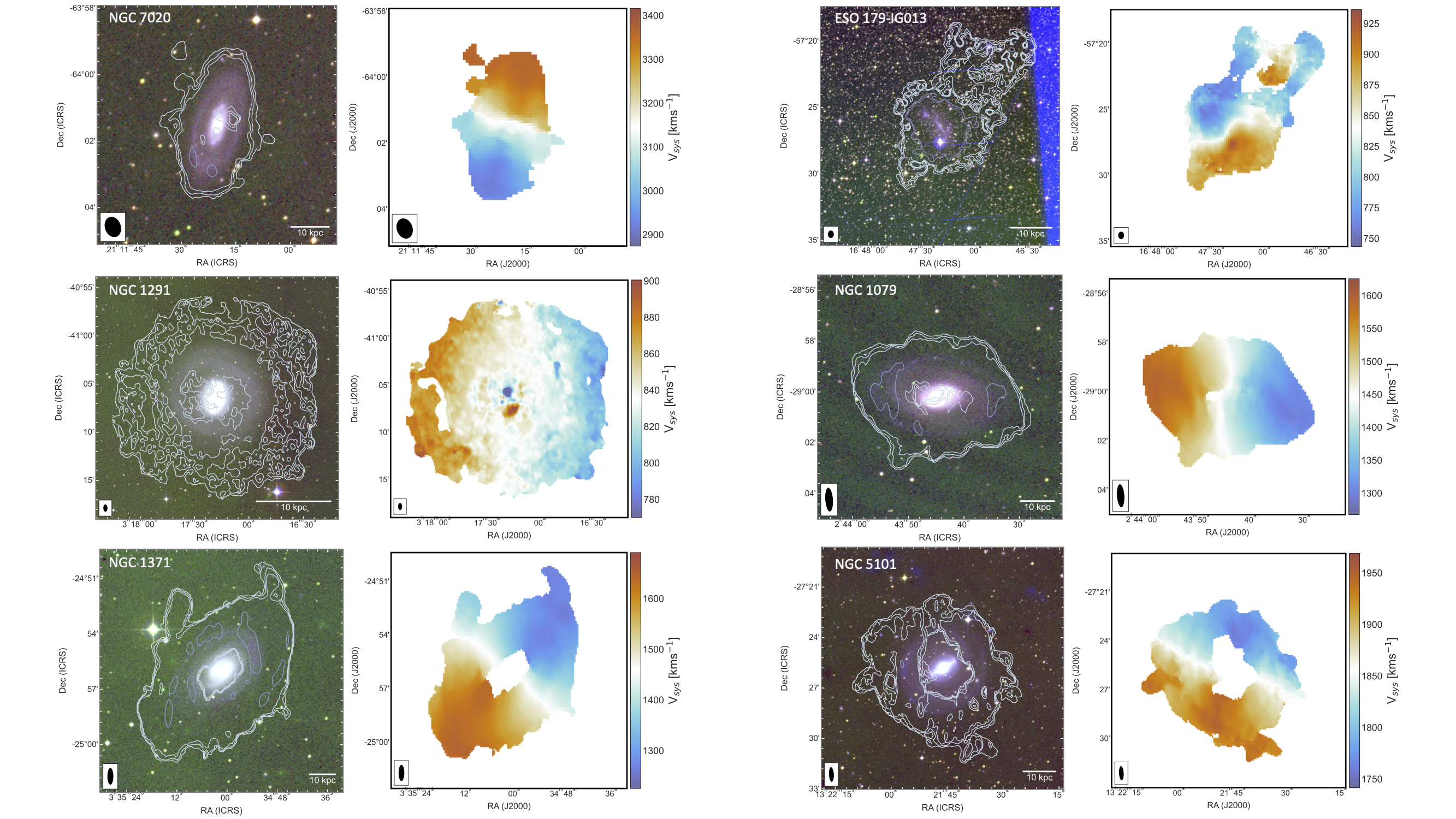}
    \caption{Same as Fig.~\ref{fig:mosaic1}}
    \label{fig:mosaic2}
    \end{figure}
\end{landscape}

\begin{landscape}
    \begin{figure}
    \centering
    \includegraphics[width=24cm,height=14cm]{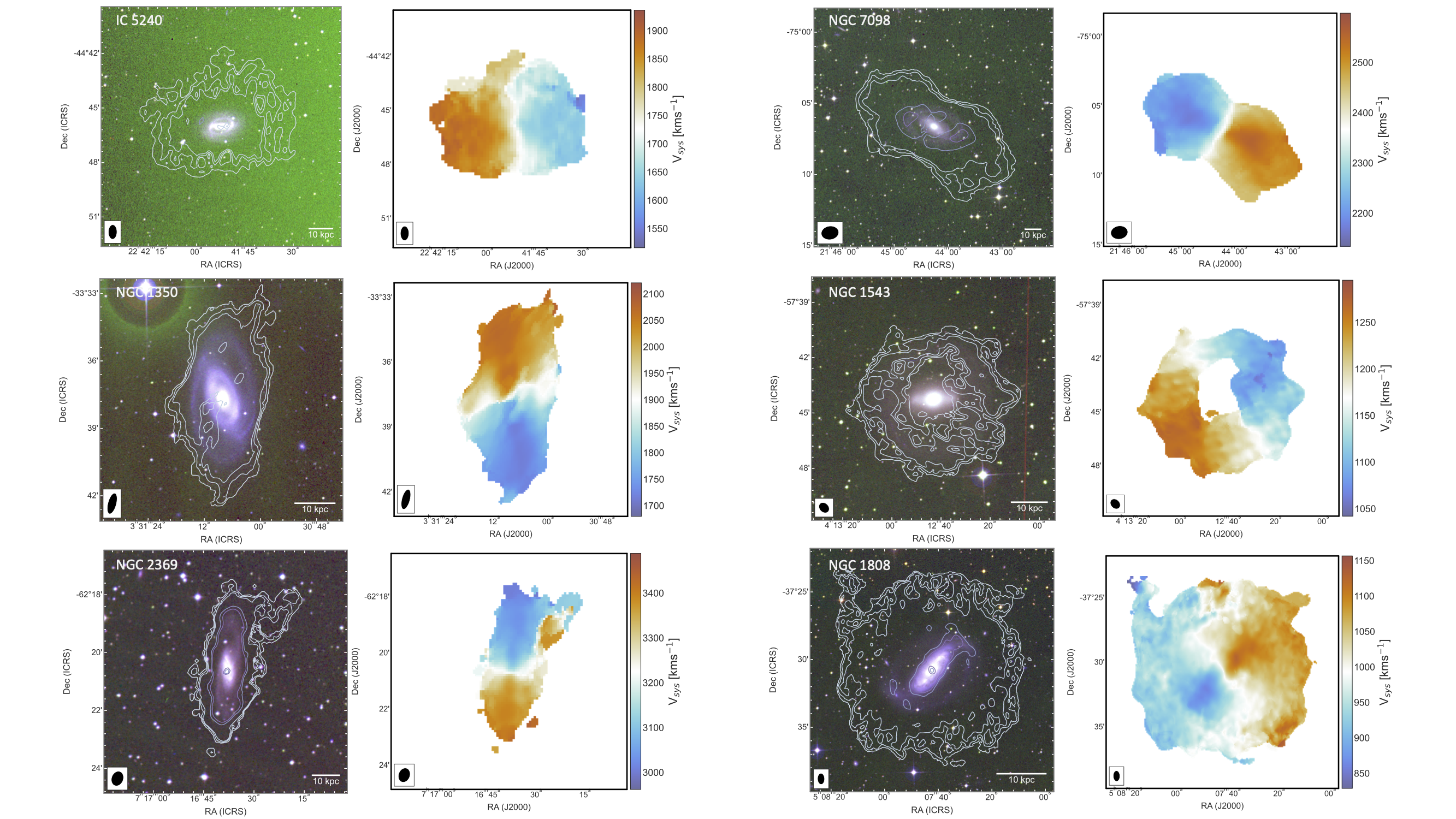}
    \caption{Same as Fig.~\ref{fig:mosaic1}}
    \label{fig:mosaic3}
    \end{figure}
\end{landscape}

\begin{landscape}
    \begin{figure}
    \centering
    \includegraphics[width=24cm,height=14cm]{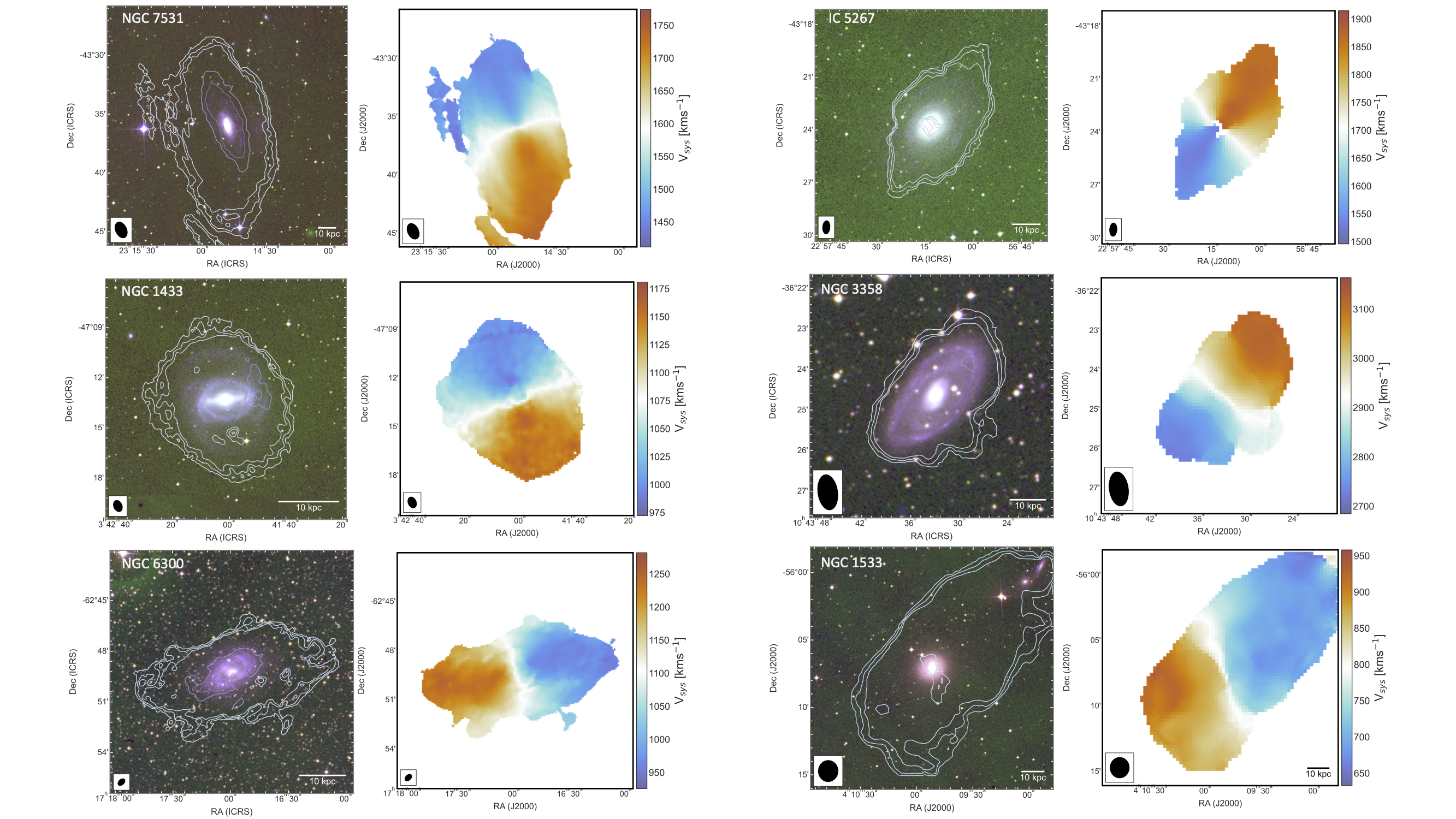}
    \caption{Same as Fig.~\ref{fig:mosaic1}}
    \label{fig:mosaic4}
    \end{figure}
\end{landscape}

\begin{figure*}
    \centering
    \includegraphics[width=0.87\columnwidth]{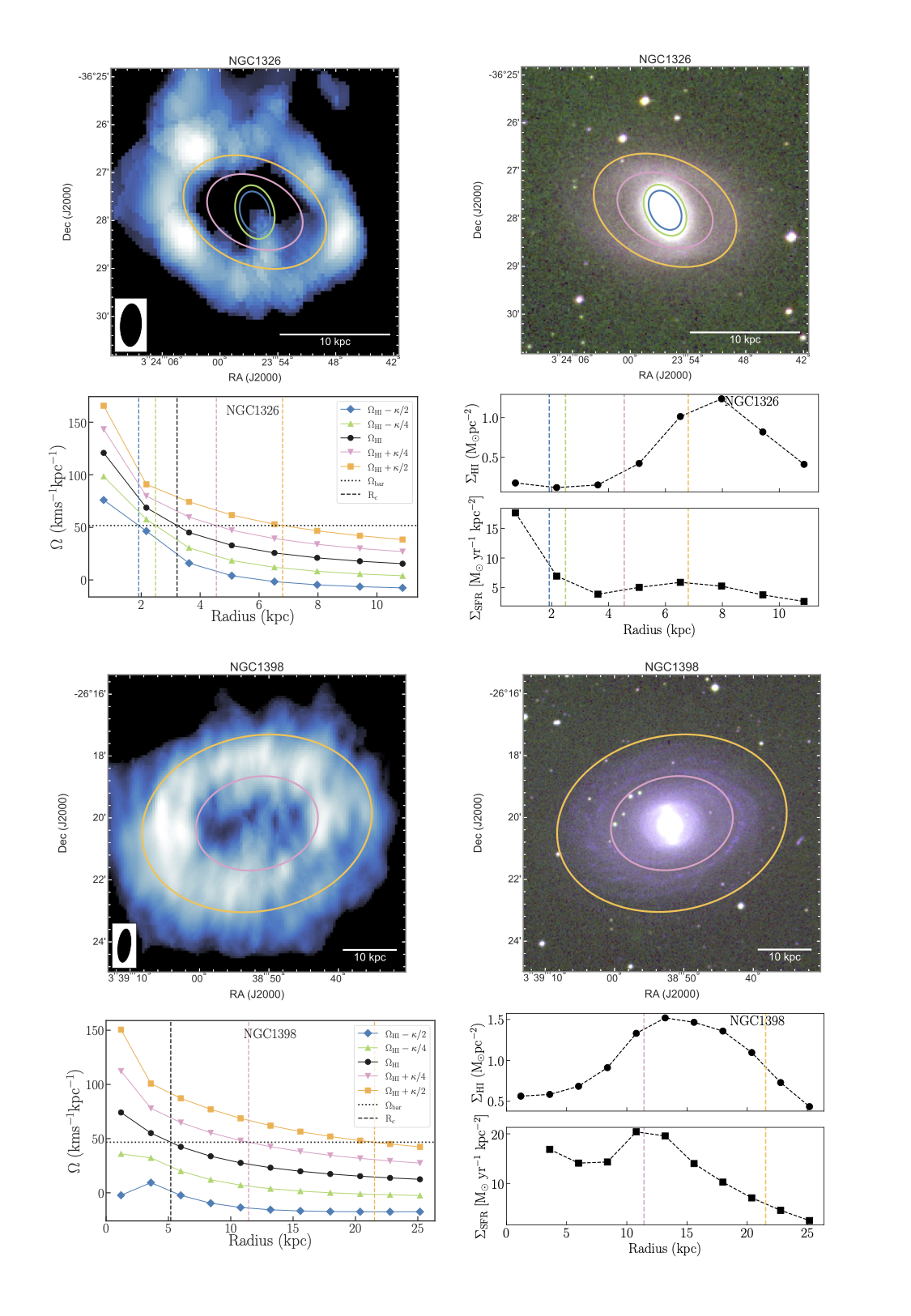}
    \caption{\textit{Top row}: Image to the left shows the moment 0 \h1 intensity map and image to the right shows the composite DSS image of the galaxy NGC 1326. \textit{Second row}: Left panel shows the angular velocity ($\Omega_{\textrm{\h1}}$) profile in black, while the blue, green, pink and orange profiles denote the various Lindblad curves. The radius of co-rotation ($R_c$) is denoted by the black vertical dashed line. The blue, green, pink and orange ellipses (and vertical dashed lines) always represent the ILR, iUHR, oUHR and OLR locations. The bottom two rows are similar to the first two rows but shown for the galaxy NGC 1398.}
    \label{fig:resonance_rings_appendix1}
\end{figure*}

\begin{figure*}
    \centering
    \includegraphics[width=0.87\columnwidth]{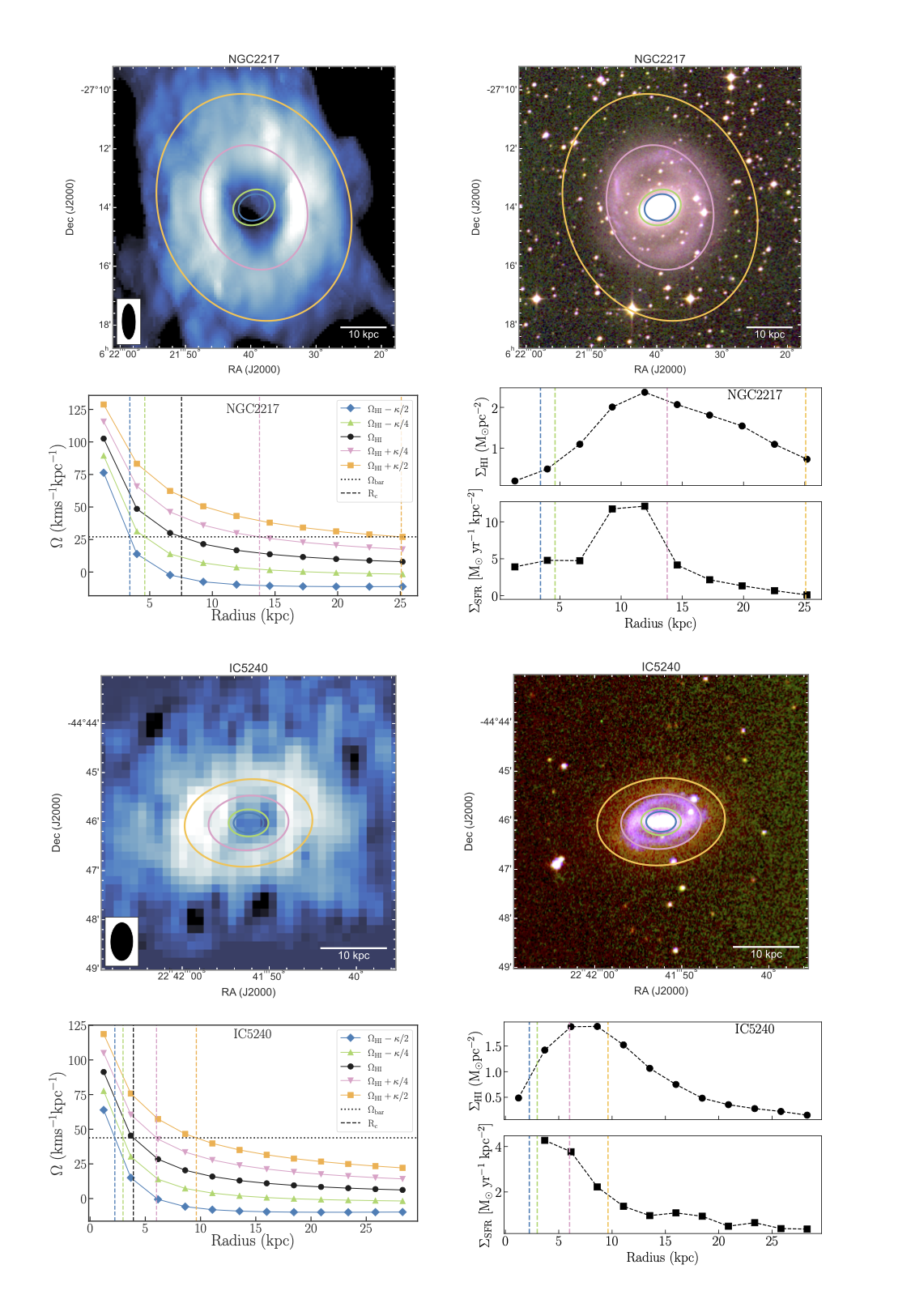}
    \caption{\textit{Top row}: Image to the left shows the moment 0 \h1 intensity map and image to the right shows the composite DSS image of the galaxy NGC 2217. \textit{Second row}: Left panel shows the angular velocity ($\Omega_{\textrm{\h1}}$) profile in black, while the blue, green, pink and orange profiles denote the various Lindblad curves. The radius of co-rotation ($R_c$) is denoted by the black vertical dashed line. The blue, green, pink and orange ellipses (and vertical dashed lines) always represent the ILR, iUHR, oUHR and OLR locations. The bottom two rows are similar to the first two rows but shown for the galaxy IC 5240.}
    \label{fig:resonance_rings_appendix2}
\end{figure*}

\begin{figure*}
    \centering
    \includegraphics[width=0.87\columnwidth]{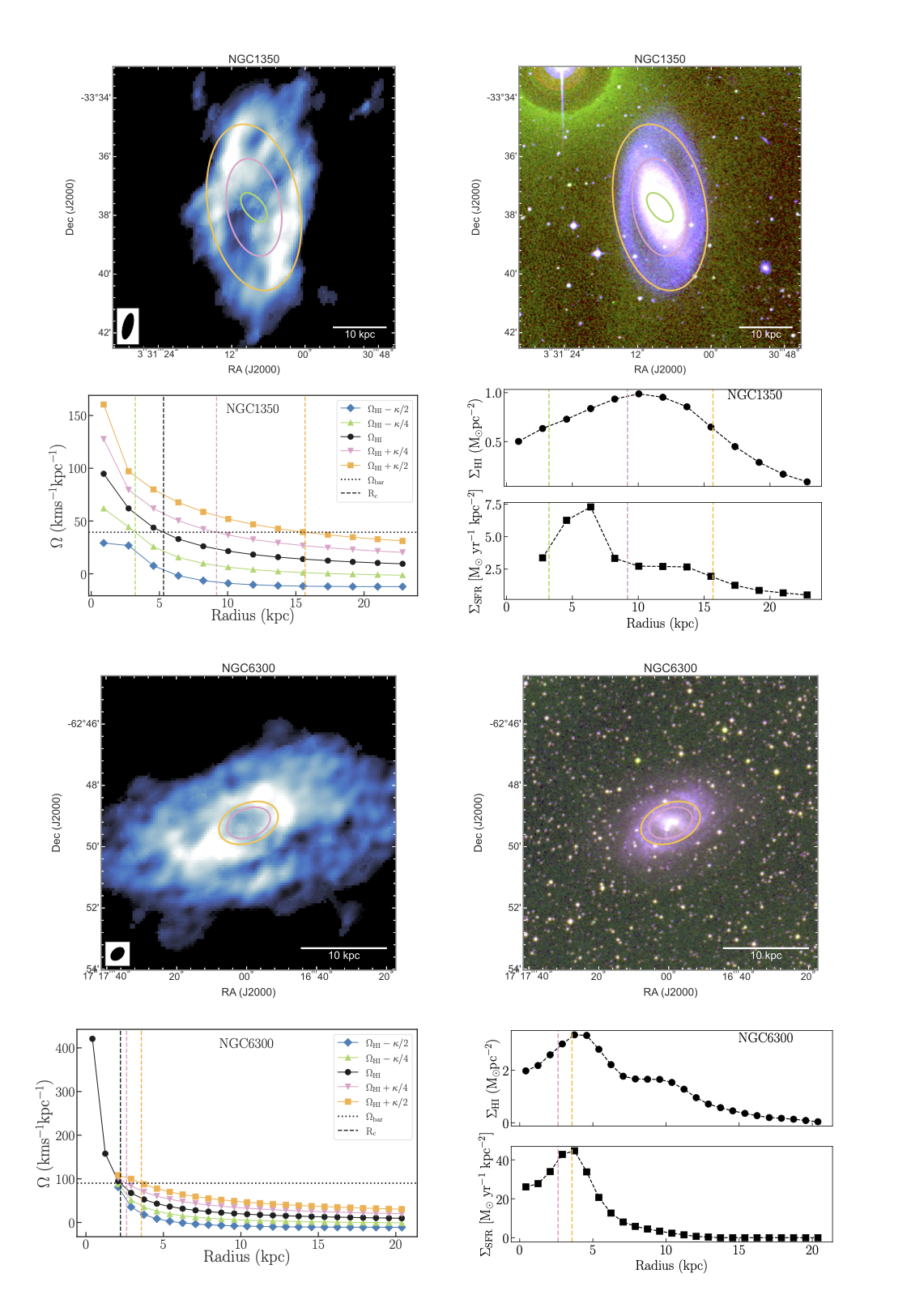}
    \caption{\textit{Top row}: Image to the left shows the moment 0 \h1 intensity map and image to the right shows the composite DSS image of the galaxy NGC 1350. \textit{Second row}: Left panel shows the angular velocity ($\Omega_{\textrm{\h1}}$) profile in black, while the blue, green, pink and orange profiles denote the various Lindblad curves. The radius of co-rotation ($R_c$) is denoted by the black vertical dashed line. The blue, green, pink and orange ellipses (and vertical dashed lines) always represent the ILR, iUHR, oUHR and OLR locations. The bottom two rows are similar to the first two rows but shown for the galaxy NGC 6300.}
    \label{fig:resonance_rings_appendix3}
\end{figure*}

\begin{figure*}
    \centering
    \includegraphics[width=0.87\columnwidth]{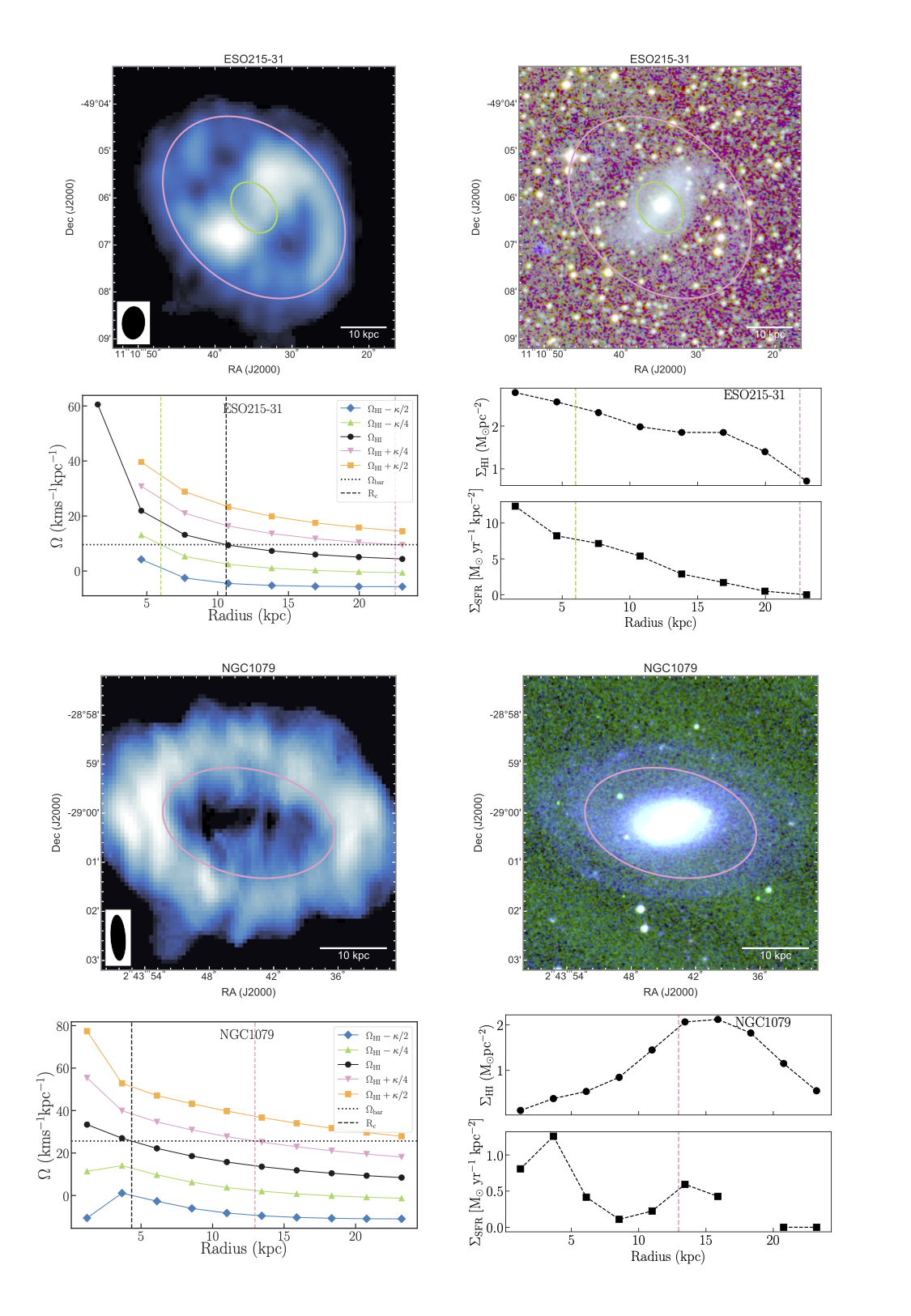}
    \caption{\textit{Top row}: Image to the left shows the moment 0 \h1 intensity map and image to the right shows the composite DSS image of the galaxy ESO 215-31. \textit{Second row}: Left panel shows the angular velocity ($\Omega_{\textrm{\h1}}$) profile in black, while the blue, green, pink and orange profiles denote the various Lindblad curves. The radius of co-rotation ($R_c$) is denoted by the black vertical dashed line. The blue, green, pink and orange ellipses (and vertical dashed lines) always represent the ILR, iUHR, oUHR and OLR locations. The bottom two rows are similar to the first two rows but shown for the galaxy NGC 1079.}
    \label{fig:resonance_rings_appendix4}
\end{figure*}

\begin{figure*}
    \centering
    \includegraphics[width=0.87\columnwidth]{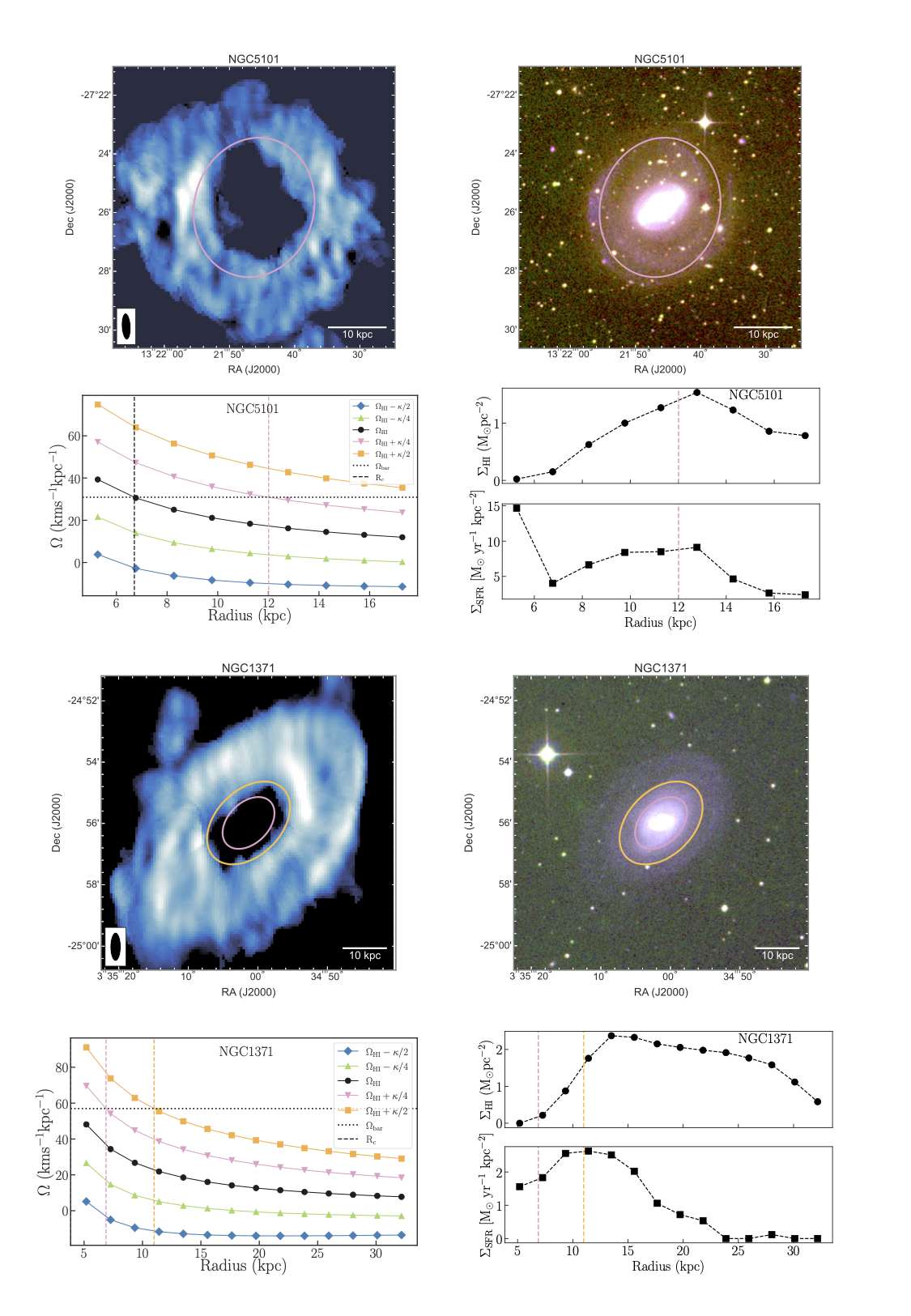}
    \caption{\textit{Top row}: Image to the left shows the moment 0 \h1 intensity map and image to the right shows the composite DSS image of the galaxy NGC 5101. \textit{Second row}: Left panel shows the angular velocity ($\Omega_{\textrm{\h1}}$) profile in black, while the blue, green, pink and orange profiles denote the various Lindblad curves. The radius of co-rotation ($R_c$) is denoted by the black vertical dashed line. The blue, green, pink and orange ellipses (and vertical dashed lines) always represent the ILR, iUHR, oUHR and OLR locations. The bottom two rows are similar to the first two rows but shown for the galaxy NGC 1371.}
    \label{fig:resonance_rings_appendix5}
\end{figure*}

\begin{landscape}
    \begin{figure}
    \centering
    \includegraphics[width=24cm,height=14cm]{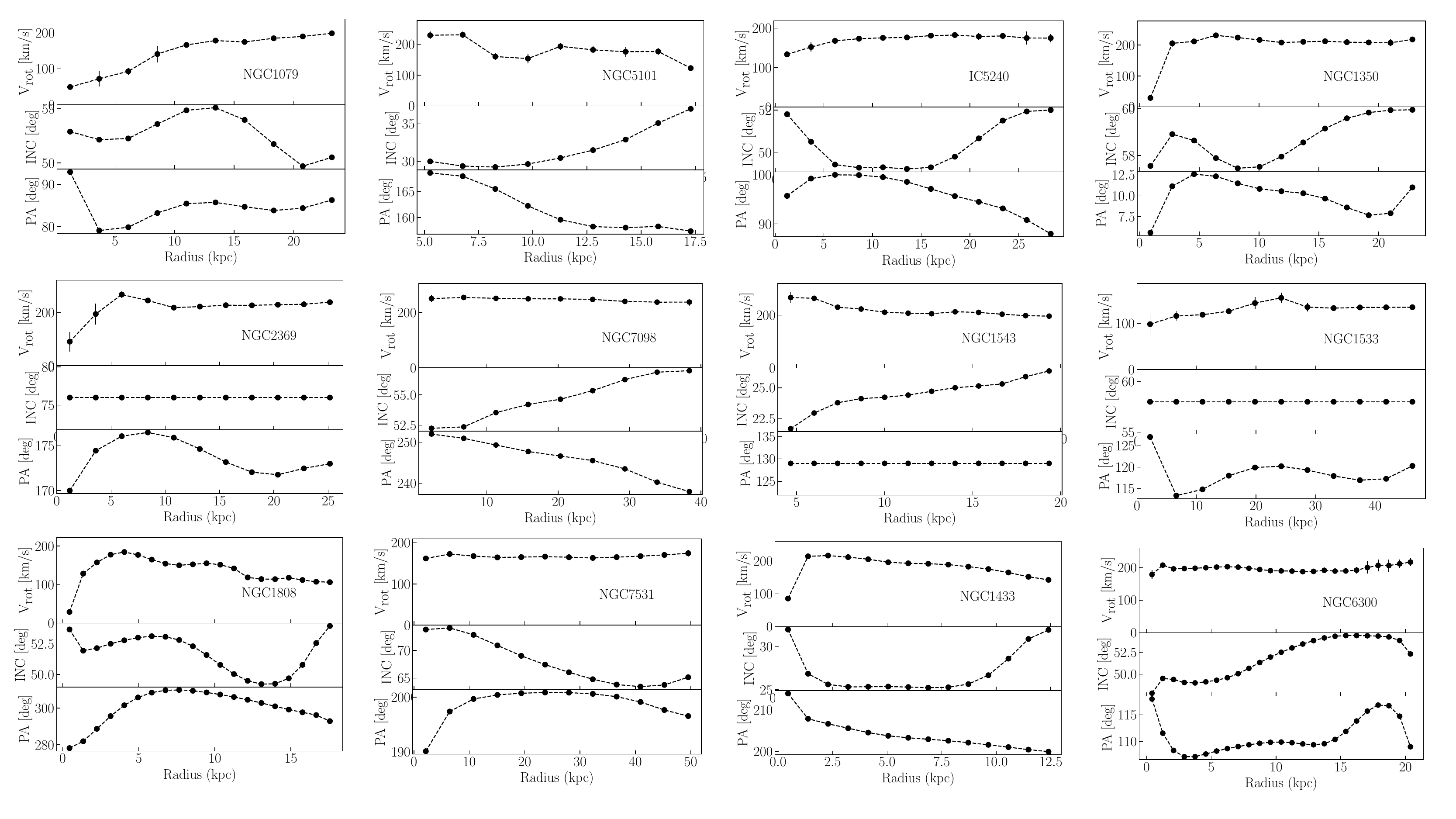}
    \caption{Each panel shows the plots of the rotation curve (v$_{\textrm{rot}}$) and the radial variation of the inclination ($i$) and position angles (PA) from the 3DBarolo fits to the galaxies.}
    \label{fig:parameter_plots1}
    \end{figure}
\end{landscape}

\begin{landscape}
    \begin{figure}
    \centering
    \includegraphics[width=24cm,height=14cm]{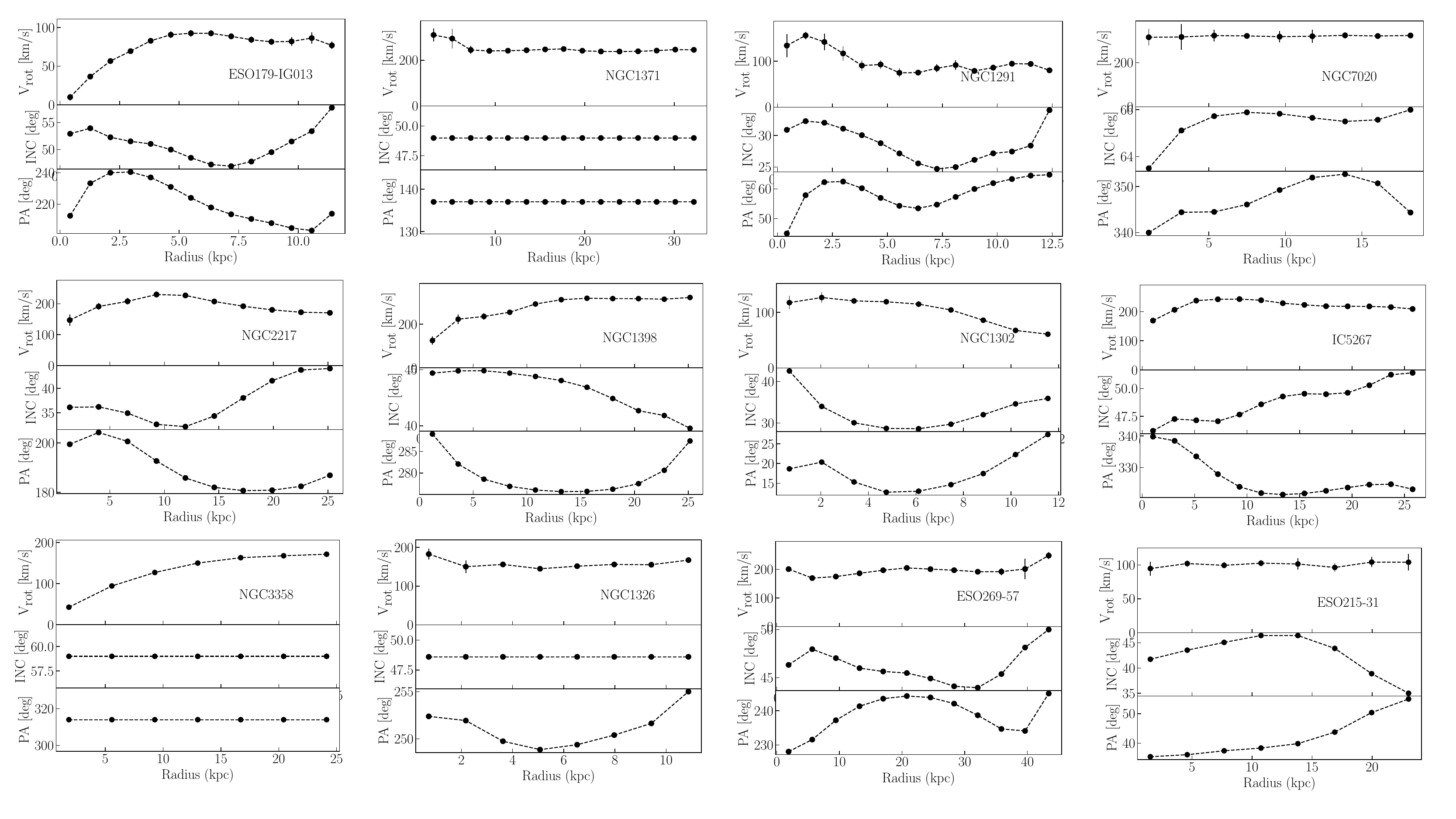}
    \caption{Each panel shows the plots of the rotation curve (v$_{\textrm{rot}}$) and the radial variation of the inclination ($i$) and position angles (PA) from the 3DBarolo fits to the galaxies.}
    \label{fig:parameter_plots2}
    \end{figure}
\end{landscape}

\begin{figure*}
\centering
\includegraphics[height=22cm,width=18cm]{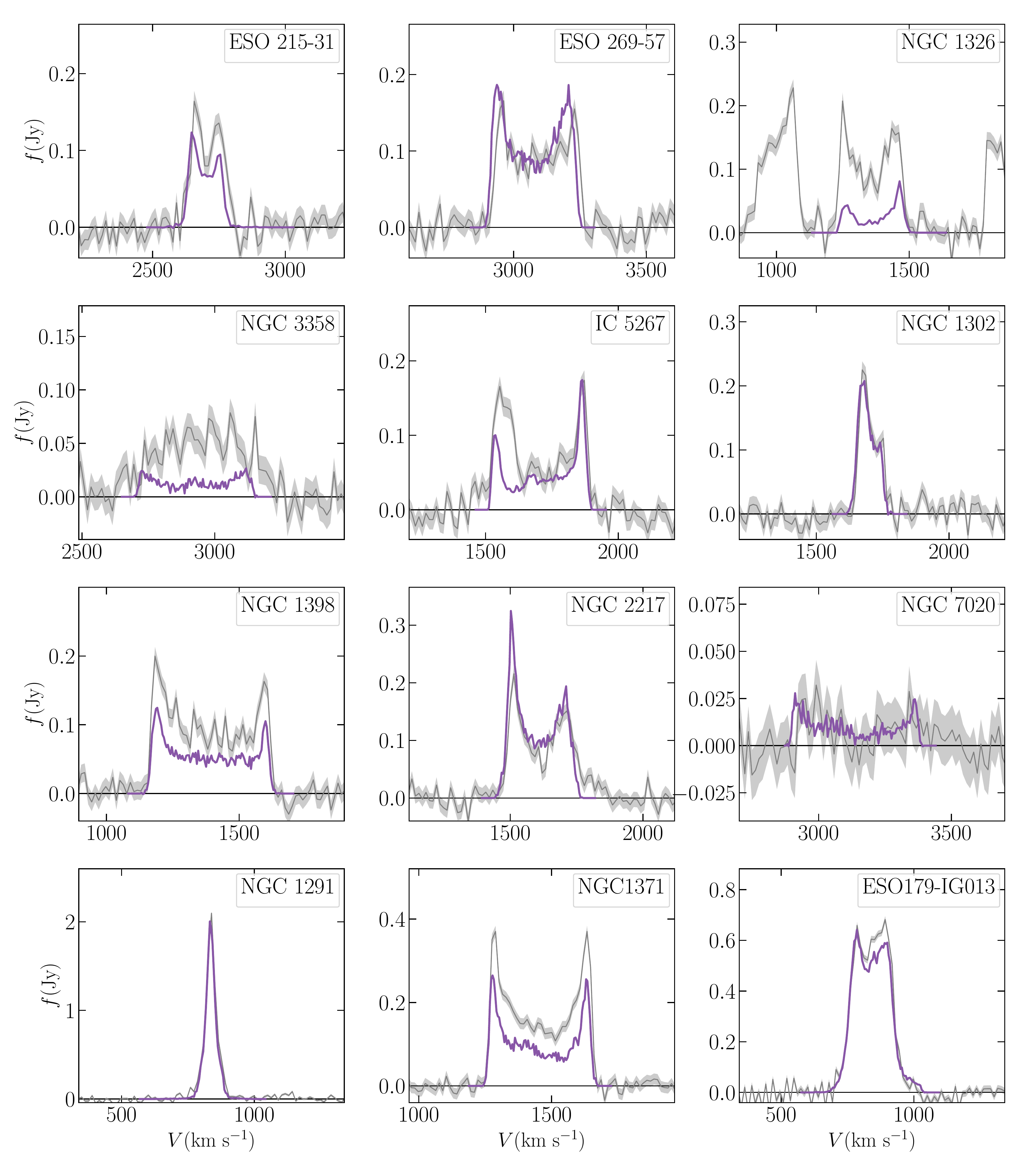}
\caption{The HI spectra for the \h1-RINGS sample. The grey lines are spectra from HIPASS with shaded rms of 13.3mJy and the purple lines are spectra from ATCA observations. }
\label{fig:spectra_1}
\end{figure*}

\begin{figure*}
\centering
\includegraphics[height=22cm,width=18cm]{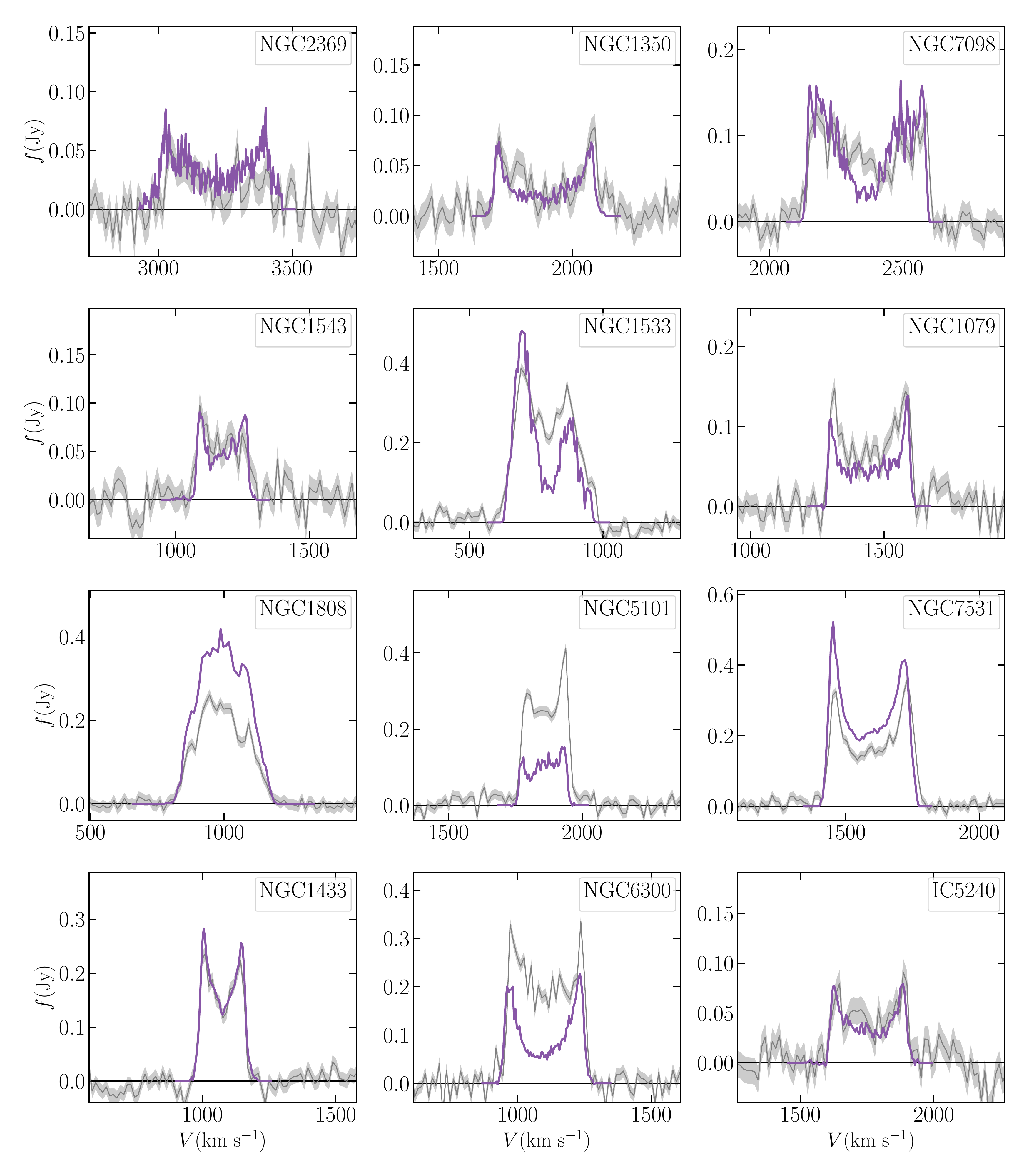}
\caption{The HI spectra for the \h1-RINGS sample. The grey lines are spectra from HIPASS with shaded rms of 13.3mJy and the purple lines are spectra from ATCA observations.}
\label{fig:spectra_2}
\end{figure*}

\end{document}